\documentclass{aa}  
\usepackage{graphicx}
\usepackage{txfonts}
\def \IZ{I$\,$Zw$\,$18}

\def \LX{\mbox{$L_{\rm X} $}}
\def \TX{\mbox{$T_{\rm X} $}}
\def \MdX{\mbox{$M2_{\rm X} $}}
\def \MdX2{\mbox{$M2_{\rm X2} $}}

\def \Msun{\mbox{\,$M_{\odot} $}}

\def \Ne{\mbox{$N_{\rm e} $}}
\def \Nh{\mbox{$N_{\rm H} $}}
\def \Te{\mbox{$T_{\rm e} $}}
\def \arcsec{\mbox{$^{\prime\prime}$}}

\def \lya{Ly$\alpha$}
\def \ha{H$\alpha$}
\def \hb{H$\beta$}

\def \hi{\ion{H}{i}}
\def \hii{\ion{H}{ii}}
\def \hei{\ion{He}{i}}
\def \heii{\ion{He}{ii}}
\def \cii{[\ion{C}{ii}]}
\def \ciii{\ion{C}{iii}]}
\def \civ{\ion{C}{iv}}
\def \nii{[\ion{N}{ii}]}
\def \oi{[\ion{O}{i}]}
\def \oii{[\ion{O}{ii}]}
\def \oiii{[\ion{O}{iii}]}
\def \oiiis{\ion{O}{iii}]}
\def \oiv{[\ion{O}{iv}]}
\def \oivs{\ion{O}{iv}]}
\def \neii{[\ion{Ne}{ii}]}
\def \neiii{[\ion{Ne}{iii}]}
\def \neiv{[\ion{Ne}{iv}]}
\def \mgi{\ion{Mg}{i}]}
\def \aliii{\ion{Al}{iii}}
\def \sid{[\ion{Si}{ii}]}
\def \sit{\ion{Si}{iii}]}
\def \siq{\ion{Si}{iv}}
\def \sii{[\ion{S}{ii}]}
\def \siii{[\ion{S}{iii}]}
\def \siv{[\ion{S}{iv}]}
\def \arii{[\ion{Ar}{ii}]}
\def \ariii{[\ion{Ar}{iii}]}
\def \ariv{[\ion{Ar}{iv}]}
\def \arv{[\ion{Ar}{v}]}
\def \feii{[\ion{Fe}{ii}]}
\def \feiii{[\ion{Fe}{iii}]}
\def \feiv{[\ion{Fe}{iv}]}
\def \fev{[\ion{Fe}{v}]}
\def \fevi{[\ion{Fe}{vi}]}
\def \la{$\lambda$\,}

\def \ie{{\it i.e.}}
\def \eg{{\it e.g.}}

\def \ergs{\,erg\,s$^{-1}$}
\def \ergcs{\,erg\,cm$^{-2}$\,s$^{-1}$}

\def \My{\,Myrs}

\def \dx{\,dex}
\def \cm2{\,cm$^2$}
\def \cc{\,cm$^{-3}$}
\begin{document}
  \title{Heating of blue compact dwarf galaxies: 
gas distribution and photoionization by stars in I\,Zw 18}

%%%   \subtitle{I. Overviewing the $\kappa$-mechanism}

  \author{D. P\'equignot
          }

   \offprints{D. P\'equignot}

   \institute{LUTH, Observatoire de Paris, CNRS, Universit\'e Paris Diderot; 5 Place Jules Janssen, 92190 Meudon, France.
             }

%%LUTH, Observatoire de Paris, CNRS, Université Paris Diderot ; 5 Place Jules Janssen, 92190 Meudon, France

%%LUTH, Laboratoire l'Univers et ses Th\'eories,  
%%              associ\'e au CNRS (FRE 2462) et \`a l'Universit\'e Paris 7, 
%%              Observatoire de Paris-Meudon, F-92195 
%%              Meudon C\'edex, France.

   \date{Received ? / Accepted ?}

   \abstract
{}
{Photoionization models so far are unable to account for the high electron 
temperature $T_e$([\ion{O}{iii}]) implied by the line intensity ratio 
[\ion{O}{iii}]$\lambda$4363\AA/[\ion{O}{iii}]$\lambda$5007\AA\ in 
low-metallicity blue compact dwarf galaxies, casting doubts on the 
assumption of photoionization by hot stars as the dominant source of heating 
of the gas in these objects of large cosmological significance.}
{Combinations 
of runs of the 1-D photoionization code NEBU are used to explore 
alternative models for the prototype giant \ion{H}{ii} region shell 
I\,Zw\,18\,NW, with no reference to the filling factor concept and with 
due consideration for geometrical and stellar evolution constraints.}
{Acceptable models for I\,Zw\,18\,NW are obtained, which represent 
schematically an incomplete shell comprising radiation-bounded condensations 
embedded in a low-density matter-bounded diffuse medium. The thermal 
pressure contrast between gas components is about a factor 7. 
The diffuse phase can be in pressure balance with the hot superbubble 
fed by mechanical energy from the inner massive star cluster. The 
failure of previous modellings is ascribed to (1) the adoption 
of an inadequate small-scale gas density distribution, which proves 
critical when the collisional excitation of hydrogen contributes 
significantly to the cooling of the gas, and possibly (2) a too 
restrictive implementation of Wolf-Rayet stars in synthetic stellar 
cluster spectral energy distributions. A neutral gas component heated 
by soft X-rays, whose power is less than 1\% of the star cluster 
luminosity and consistent with CHANDRA data, can explain 
the low-ionization fine-structure lines detected by SPITZER. 
[O/Fe] is slightly smaller in I\,Zw\,18\,NW than in Galactic Halo 
stars of similar metallicity and [C/O] is correlatively large.}
{Extra heating by, \eg, dissipation of mechanical energy is not 
required to explain $T_e$([\ion{O}{iii}]) in I\,Zw\,18. Important 
astrophysical developments are at stakes in the 5\% uncertainty attached 
to \oiii\ collision strengths.} 


 %%   \authorrunning{P\'equignot}

 %%   \titlerunning{Heating of I\,Zw\,18}

 %%   %\headnote{Letter to the Editor}

   \maketitle
%
%________________________________________________________________

\section{Introduction}
\label{intro}

The optical properties of Blue Compact Dwarf (BCD) galaxies are 
similar to those of Giant Extragalactic \hii\ Regions (GEHIIR). Their 
blue continuum arises from one or several young Massive Star Clusters (MSC), 
which harbour extremely large numbers of massive stars. 

BCDs are relatively isolated, small-sized, metal-poor galaxies 
(Kunth \& \"Ostlin 2000) and may be the rare `living fossils' of a formerly 
common population. BCDs can provide invaluable pieces of information 
about the primordial abundance of helium (\eg, Davidson \& Kinman 1985), 
the chemical composition of the InterStellar Medium 
(ISM, \eg, Izotov et al. 2006), the formation and evolution of 
massive stars, and the early evolution of galaxies at large redshift. 
Among them, \IZ\ stands out as one of the most oxygen-poor BCDs known 
(\eg, Izotov et al. 1999) and a young galaxy candidate in 
the Local Universe (\eg, Izotov \& Thuan 2004). 

The line emission of \hii\ regions is believed to be governed by 
radiation from massive stars, but spectroscopic diagnostics most 
often indicate spatial fluctuations of the electron temperature \Te\ 
(see the dimensionless parameter $t^2$, Peimbert, 1967), that appear 
larger than those computed in {\sl usual} photoionization models, 
suggesting an {\sl extra heating} of the emitting gas
(\eg, Peimbert, 1995; Luridiana et al. 1999). Until the 
cause(s) of this failure of photoionization models can be identified, 
a sword of Damocles is hung over a basic tool of astrophysics. 

Tsamis \& P\'equignot (2005) showed that, in the GEHIIR 30\,Dor of the LMC, 
the various \Te\ diagnostics could be made compatible with one another if the 
ionized gas were {\sl chemically inhomogeneous} over small spatial scales. 
A pure photoionization model could then account for the spectrum of a bright 
filament of this nebula. Although this new model needs confirmation, it is 
in suggestive agreement with a scenario by Tenorio-Tagle (1996) of a recycling 
of supernova ejecta through a rain of metal-rich droplets cooling and 
condensing in the Galaxy halo, then falling back on to the Galactic disc 
and incorporating into the ISM without significant mixing until a new \hii\ 
region eventually forms. If this class of photoionization models is finally 
accepted, extra heating will not be required for objects like 30\,Dor, 
with near Galactic metallicity. 

Another problem is encountered in low-metallicity (`low-Z') BCDs (Appendix~A). 
In BCDs, available spectroscopic data do not provide signatures for 
$t^2$'s, but a major concern of photoionization models is explaining the 
high temperature \Te(\oiii) infered from the observed intensity 
ratio $r$(\oiii) = \oiii\la4363/(\oiii\la5007+4959). Thus, 
Stasi\'nska \& Schaerer (1999, SS99) conclude that photoionization 
by stars fails to explain $r$(\oiii) in the GEHIIR \IZ\,NW and that 
photoionization must be supplemented by other heating mechanisms. 
A requirement for extra heating is indirectly stated by 
Luridiana et al. (1999) for NGC\,2363. 

A possible heating mechanism is conversion of mechanical energy 
provided by stellar winds and supernovae, although a conclusion 
of Luridiana et al. (2001) does not invite to optimism. 
A limitation of this mechanism is that most of 
this mechanical energy is likely to dissipate in hot, steadily 
expanding superbubbles (Martin, 1996; Tenorio-Tagle et al. 2006). 
It is doubtful that heat conduction from this coronal gas could 
induce enough localized enhancement of \Te\ in the photoionized gas 
(\eg, Maciejewski et al. 1996), even though Slavin et al. (1993) 
suggest that turbulent mixing may favour an energy transfer. 
Martin (1997) suggests that shocks could help to explain the trend 
of ionization throughout the diffuse interstellar gas of BCDs, but 
concedes that ``shocks are only being invoked as a secondary 
signal in gas with very low surface brightness''. 
Finally, photoelectric heating from dust is inefficient in metal-poor 
hot gas conditions (Bakes \& Tielens 1994). 

Nevertheless, the conclusion of SS99 is now accepted in many studies 
of GEHIIRs. It entails so far-reaching consequences concerning the 
physics of galaxies at large redshifts as to deserve close scrutiny. If, 
for exemple, the difference between observed and computed \Te(\oiii) 
in the model by SS99 were to be accounted for by artificially 
raising the heat input proportionally to the photoionization heating, 
then the total heat input in the emitting gas should be doubled. {\sl This 
problem therefore deals with the global energetics of the early universe}. 

After reviewing previous models for \IZ\,NW (Sect.~\ref{prev_IZ}), 
observations and new photoionization models are described in Sects.~\ref{obs} 
\& \ref{newmod}. Results presented in Sect.~\ref{res} are discussed in 
Sect.~\ref{disc}. Concluding remarks appear in Sect.~\ref{concl}. Models 
for other GEHIIRs are reviewed in Appendix~A. Concepts undelying the 
new photoionization models are stated in Appendices~B and C. 

\section{Photoionization models for \IZ\,NW}
\label{prev_IZ}

Early models are reviewed by SS99. Dufour et al. (1988)
envisioned a collection of small \hii\ regions of different 
excitations. Campbell (1990) proposed to enhance $r$(\oiii) by 
collisional quenching of \oiii\la5007 in an ultra-compact structure 
(electron density \Ne~=~10$^5$\cc). Stevenson et al. (1993) modelled 
a uniform sphere of radius $\sim$~0.4\arcsec. Fairly satisfactory 
computed emission-line spectra were obtained, but the model \hii\ 
regions were inacceptably compact according to subsequent imaging. 
Firstly, \IZ\,NW is essentially an incomplete \hii\ region shell 
of some 5\arcsec\ in diameter surrounding a young MSC, 
which is {\sl not spatially coincident} with the ionized gas. 
Secondly, both the highly \Te-sensitive line \oiii\la4363 and the 
high-ionization line \heii\la4686 are detected throughout the 
whole shell and beyond. 

\subsection{Recent modelling attempts}
\label{prev_recent}

According to SS99, a `model' is basically a uniform, matter-bounded spherical 
shell, whose only free parameter is a {\sl filling factor} $\epsilon$. 
The hydrogen density is \Nh\ = 10$^2$\cc, inspired by the electron 
density \Ne(\sii) derived from the observed doublet intensity ratio 
$r$(\sii) = \sii\la6716/\sii\la6731. The central ionizing source is a 
synthetic stellar cluster, which fits the observed continuum flux at 3327\AA\ 
and maximizes the nebular \heii\ emission. The inner angular radius is 
1.5\arcsec. The outer radius $r_{\rm out}$ is defined by 
the condition that the computed \oiii\la5007/\oii\la3727 ratio fits 
the observed one. For increasing $\epsilon$, the material is on average 
closer to the source and more ionized, which must be compensated for by 
increasing the optical depth to keep \oiii/\oii\ constant, so that the 
computed \hb\ flux increases and \heii/\hb\ decreases. For $\epsilon$ $>$ 0.1, 
the shell becomes radiation bounded, \oiii/\oii\ 
grows larger than it should and $r_{\rm out}$ becomes less 
than the observed value ($\sim$~2.5\arcsec). Because 
of these trends, SS99 discard large-$\epsilon$ models and select 
a model with $\epsilon$~$\sim$~0.01, on the basis that the computed 
$r_{\rm out}$, \hb\ and \heii\ are roughly acceptable\footnote{In the context, 
the \hb\ and \heii\ fluxes are poor selection criteria for $\epsilon$, 
since the former is proportional to the (unknown) covering factor of the 
shell and the latter depends much on questionable synthetic stellar cluster 
spectra (Appendix~\ref{WRstars}).}. This `best model' presents two major 
drawbacks: 
(1) as stated by SS99, the computed $r$(\oiii) is too small by a highly 
significant factor 1.3, and (2) \sii\ and \oi\ are grossly underestimated. 

\subsection{The \oiii\ line problem}
\label{prev_oiii}

Concerning $r$(\oiii), SS99 note without justification that, for 
different values of \Nh, ``no acceptable solution is found''. It will 
become clear that this cursory statement is central for concluding that 
extra heating is required. 

In response, Viegas (2002, hereafter V02) makes 
the correct point that adopting a density less than 
10$^2$\cc\ (and $\epsilon$ = 1) can help improving the computed $r$(\oiii). 
However, in the example shown by V02 (\Ne\ = 30\cc), $r$(\oiii) is still 
10\% low\footnote{The tolerance of 9\% allowed by SS99 (adopted by V02) is 
probably too large (Sect.~\ref{spectr}). This error bar 
is justified in the logics of SS99, who aim to demonstrate an 
{\sl absence} of solution.} and $r$(\sii) is somewhat off. 
Moreover, not only \sii\ and \oi, but now \oii\ {\sl as well} is 
strongly underpredicted, in accordance with the analysis of SS99. 
V02 then proposes that radiation-bounded filaments with density 
10$^4$\cc\ are embedded in the low-density gas at different distances 
from the source. If the emission of \sii\ and \oi\ (together with \oii) 
can indeed be much increased in this way, this denser component entails 
serious difficulties. Firstly, since the computed \oiii/\oii\ 
ratio is not very much less than the observed one in these filaments, 
a sizeable fraction of \oiii\ must come from them (together with \oii) 
and since, due to enhanced \hi\ cooling (Appendix\,\ref{filling}), 
$r$(\oiii) is now {\sl half} the observed value, any composite model 
accounting simultaneously for \oiii\la5007 and \oii\la3727 will 
underpredict $r$(\oiii) in the same manner as the uniform model of SS99. 
Secondly, since at least half the \sii\ emission should come from the 
filaments, in which $r$(\sii) is again only half the observed value 
(large \Ne), the composite $r$(\sii) will be inacceptably off. Thirdly, 
no explicit solution is exhibited and it is unclear how a composite 
model of the kind envisaged by V02 will simultaneouly match all lines. 
From the evidences she presents, V02 is not founded to claim 
that ``pure photoionization can explain \IZ\ observations''. 
The inconclusiveness of the alternative she proposes 
effectively reinforces the standpoint of SS99. 

\subsection{The \oi\ and \sii\ line problem}
\label{prev_oi}

If SS99 regard the \oiii\ discrepancy as a highly significant feature, 
they are optimistic concerning the photoionization origin 
of \oi\la6300+63, underpredicted by 2\dx\ in their model.
Two configurations are envisaged by SS99. 

\subsubsection{Extremely dense filaments?}
\label{prev_oidens}

In a first configuration, radiation-bounded filaments of 
density 10$^6$\cc\ are embedded in the \hii\ region: the 
density is so high as to severely quench most lines other than 
\oi\ and \hi: only $\sim$~10\% of \sii\la6716+31 arises from 
these filaments, ensuring that $r$(\sii) is not much influenced. 
This attempt to solve in anticipation the problem met by V02 
(Sect.~\ref{prev_oiii}) raises three difficulties, however: 
(1) condensations which 
contrast in density by a factor of 10$^4$ with their surroundings and 
present a large enough covering factor ($\sim$\,10\% according to SS99) 
as to intercept a significant fraction of the primary radiation 
would probably represent most of the mass; 
(2) since the main body of the model \hii\ region produces only one 
quarter of the observed \sii\la6716+31 flux, it is not clear where this 
doublet would be emitted\footnote{SS99 prefer casting 
doubts on the ionization balance of S$^+$.}; and 
(3) this highly artificial, strictly dual density distribution is not the 
schematic, first-approximation representation of some more complex reality, 
rather it is a completely essential feature of the model since 
{\sl any material at intermediate densities} would usefully emit 
plenty of \sii\la6716+31, but lead to a totally wrong $r$(\sii) 
as in the description by V02. 

\subsubsection{Extremely distant filaments?}
\label{prev_oidist}

The second configuration proposed by SS99 involves radiation bounded 
`\oi\ filaments' of density 10$^2$\cc, located at $\sim$~20\arcsec\ 
from the source. If the spectroscopic objections of 
Sects.~\ref{prev_oiii} \& \ref{prev_oidens} are 
now removed since density is moderate and ionization is 
low in the filaments, new difficulties arise, notably with {\sl geometry}: 
(1) the filaments observed at $>$~10\arcsec\ or more from the NW MSC 
of \IZ\ have such a low surface brightness as to contribute negligibly 
to the brightness of the main shell (if they were projected upon it); 
(2) the spectrum of this weak emission up to $\sim$~15\arcsec\ 
$-$ ``Halo'' of V\'{\i}lchez \& Iglesias-P\'aramo (1998), 
``\ha\ Arc'' and ``Loop'' of Izotov et al. (2001) $-$ shows 
a flux ratio \oiii\la5007/\oii\la3727 of order unity, 
whereas this ratio is 1/300 in the putative \oi\ filaments, suggesting 
that the bulk of the emission observed at these distances arises from a 
gas, whose density is much less than 10$^2$\cc; 
(3) accepting all the same the existence of distant \oi\ emitting regions, 
a very peculiar geometry would be required to project these regions 
{\sl precisely and uniquely} upon the material of the irregular 
bright NW \hii\ shell to be modelled; and 
(4) in projection, this shell appears as a 1.5$-$2.5\arcsec\ 
`ring', which intercepts 1/200 of a 20\arcsec-radius sphere 
(including both the front and rear sides), incommensurable 
with the covering factor $\sim$~1/10 assumed by SS99. 

\subsection{Previous models: conclusion}
\label{prev_IZ_concl}

Attempts to model \IZ\ fail to explain not only $r$(\oiii) 
but the \oi\ and \sii\ lines as well. It is difficult to 
follow SS99 when they claim that they are ``not too far from a 
completely satisfactory photoionization model'' of \IZ. The explanatory 
value of their description is so loose as to jeopardize any inference 
drawn from it, including the requirement for extra heating in \IZ. 

In Appendix\,\ref{prevG}, a review of models obtained for other GEHIIRs 
reveals general trends and problems, which can be valuably analysed using 
the example of \IZ. 

\section{Observations of \IZ}
\label{obs}

\subsection{Basic properties}
\label{basic}

Two bright regions 5\arcsec\ apart, \IZ\,NW and SE, correspond to two 
young MSCs associated with two distinct GEHIIRs, surrounded by a common 
irregular, filamentary halo of diffuse ionized gas (\eg, Izotov et al. 2001), 
immersed in a radio \hi\,21\,cm envelope rotating around the centre of mass 
located in between the GEHIIRs (\eg, van Zee et al. 1998). Although the 
\hi\ column density peaks in the central region, large \hi\ structures 
have no stellar counterparts. A fainter cluster, 
`Component C', deprived of massive stars (no prominent \hii\ region), 
appears at 22\arcsec\ to the NW of the main body. The two young MSCs, 
1$-$5\,Myrs old, are the recent manifestations of a larger starburst, 
which started some 15\,Myrs ago in Component C and 20\,Myrs ago in the 
central region (Izotov \& Thuan 2004, IT04). A 20-25\,Myrs age is consistent 
with the dynamics of the superbubble studied by Martin (1996). In a radio 
study, Hirashita \& Hunt (2006) suggest 12$-$15\,Myrs. \IZ\ is classified as 
a `passive BCD' (\eg, Hirashita \& Hunt 2006), that is, the MSCs themselves 
are relatively diffuse, the stellar formation rate (SFR) is relatively low 
(Sect.~\ref{disc_gas}) and the starburst is not instantaneous. 

That a background population 300$-$500\,Myrs old may be the first generation 
of stars in this galaxy (Papaderos et al. 2002; IT04) is 
contested by Aloisi et al. (2007). The extended optical halo of \IZ\ is 
mostly due to ionized gas emission. Unlike for usual BCDs, the bulk of 
the stars in \IZ\ is highly concentrated, suggesting perhaps a young 
structure (Papaderos et al. 2002). 
The distance to \IZ, first quoted as 10\,Mpc, has been revised to 
$\sim$~13\,Mpc (\"Ostlin 2000) after correcting the Hubble flow 
for the attraction of the Virgo cluster. From AGB star magnitudes, 
IT04 obtain 14$\pm$1.5\,Mpc. 
At the distance $D$ = 4.0$\times$10$^{25}$\,cm (12.97\,Mpc) adopted here, 
the diameter of the bright region \IZ\,NW (5\arcsec) is over 300\,pc. 
From new deep {\sc hst} photometry revealing a red giant branch and Cepheid 
variables, Aloisi et al. (2007) obtain 18$\pm$2\,Mpc. Except for 
scaling, present results are just  marginally changed 
if this larger distance is confirmed (Sect.~\ref{disc_eva}). 

\subsection{Absolute \hi\ line fluxes and reddening}
\label{abs_red}

According to Cannon et al. (2002, CSGD02), the absolute \hb\ fluxes 
in the 5 polygons paving the NW region and the 7 polygons paving the 
SE region are 4.9 and 1.7 respectively in units of 10$^{-14}$\ergcs. 
Polygons NW D6 and SE D8 do not exactly belong to the main body of the 
\hii\ regions and are dismissed. Tenuous emission around the polygons 
is also neglected. 

The excess over the Case B recombination value of the observed average 
\ha/\hb\ ratios, 2.94 and 2.97 in the NW and SE respectively, is attributed 
to dust reddening by CSGD02, who rightly doubt the large \hi\ collisional 
excitation obtained by SS99 (Appendix\,\ref{fiat_h}). It remains that 
the Balmer decrement is influenced by collisions and that the reddening 
correction to the observed spectrum of \IZ\ has been overestimated. 
Collisional excitation results from a subtle anticorrelation between 
\Te\ and N(H$^0$)/N(H$^+$) within the nebula and can only be determined 
from a photoionization model (Contrary to a statement by CSGD02, 
the maximum effect does {\sl not} correspond to the hottest gas). 
The usually adopted recombination ratio is \ha/\hb\ = 2.75$\pm$0.01 
(Izotov et al. 1999). It is anticipated that, according to present 
models (Sect.~\ref{res}), a better \ha/\hb\ is 2.83$\pm$0.02. 
For use in the present study, published dereddened intensities 
(also corrected for stellar absorption lines) have been re-reddened 
by $\Delta$E(B-V) = $-$0.04 (in view of final results, a more 
nearly accurate correction could be $\Delta$E(B-V) = $-$0.03). 
Then the typical E(B-V) for \IZ\,NW shifts from 0.08 to 0.04, out of which 
the foreground Galactic contribution is about 0.02 (Schlegel et al. 1998). 
The reddening corrected \hb\ fluxes for the main NW and SE \hii\ regions 
are I(\hb) = 5.6 and 2.0 respectively in units of 10$^{-14}$\ergcs. 

In these units, the \ha\ flux is 33.0 over the central 
13.7$\times$10.5\arcsec\ {\sc hst} field (Hunter \& Thronson 1995; 
CSGD02) and 42.0 over a 60$\times$60\arcsec\ field (Dufour \& Hester 1990). 
Adopting overall averages \ha/\hb\ = 2.8 and E(B-V) = 0.06, 
the total dereddened \hb\ flux for \IZ\ is 18.3. 
Assuming that all of the ionizing photon sources belong to the bright 
NW and SE MSCs and that \IZ\ is globally radiation bounded, the fraction 
of photons absorbed in the two main \hii\ regions is 0.41. 
Let $Q$ and $Q_{\rm abs}$ be respectively the number of photons (s$^{-1}$) 
emitted by the MSC and absorbed by the main shell of \IZ\,NW alone. 
The fraction $Q_{\rm abs}$/$Q$ may be smaller than 0.41 for two reasons. 
Firstly, as expansion proceeds, the shells around the starbursts become 
more `porous' due to instabilities and the more evolved NW shell may 
be more affected. Assuming that no photon escape from the SE shell leads  
to a minimum $Q_{\rm abs}$/$Q$ = 0.34. A more realistic value is 
probably $Q_{\rm abs}$/$Q$ = 0.39$\pm$0.02, since a complete absorption 
in the SE would result in a strong asymmetry of the diffuse halo, 
which is not observed. Secondly, photons may escape from \IZ. 
This effect is probably weak, given the amount and extension of 
\hi\ in \IZ. The adopted nominal absorbed fraction for 
the NW shell will be $Q_{\rm abs}$/$Q$ = 0.37$\pm$0.03, with 0.30 
a conservative lower limit, obtained for a 25-30\% escape from \IZ. 

\subsection{Spectroscopic observation summary}
\label{spectr}

The optical spectrum of \IZ\ has been observed for decades 
(Sargent \& Searle 1970; Skillman \& Kennicutt, 1993, SK93; 
Legrand et al. 1997; Izotov et al. 1997, 1997b; Izotov \& Thuan, 1998; 
V\'{\i}lchez \& Iglesias-P\'aramo, 1998, VI98; Izotov et al. 1999, ICF99; 
Izotov et al. 2001; Thuan \& Izotov 2005, TI05; Izotov et al. 2006) with many 
instruments ({\sc hale}, {\sc kpno}, {\sc mmt}, {\sc keck}, {\sc cfht}, etc.), 
the UV spectrum with {\sc iue} (Dufour et al. 1988) 
and {\sc hst} (Garnett et al. 1997; Izotov \& Thuan, 1999, IT99), 
the IR spectrum with {\sc spitzer} (Wu et al. 2006, 2007), and 
the radio continuum with {\sc vla} (Hunt et al. 2005; Cannon et al. 2005). 

The {\sc iue} aperture encompasses all of the bright regions. 
In addition to \ciii\la1909, there are indications for the presence 
of \civ\la1549 and \sit\la1883+92. The {\sc hst} spectrum allows a 
direct comparison of \ciii\ with optical lines, but corresponds to 
such a limited area (0.86\arcsec) as to raise the question of the 
representativeness of the observation for \IZ\,NW as a whole. Nonetheless, 
\ciii/\hb\ is identical within 10\% in available measurements, 
once the re-evaluation of the \ha\ flux within the {\sc iue} 
aperture is taken into account (Dufour \& Hester 1990). 

The high-resolution mid-IR spectra of \IZ\ (Wu et al. 2007, Wu07) are secured 
with a 4.7$\times$11.3\arcsec\ slit. Over the 13.7$\times$10.5\arcsec\ 
{\sc hst} field, the de-redenned \hb\ flux is 13.4, while the flux from 
strictly the two central \hii\ regions, which fill only part of the 
{\sc spitzer} slit, is 7.6. The adopted \hb\ flux corresponding to the 
mid-IR spectra is taken as 10$\pm$1, the  value also used 
by Dufour \& Hester (1990) for the (partial) {\sc iue} aperture. 
Measuring line fluxes on the published tracings show excellent 
agreement with tabulated values, except for \siii\,18.7$\mu$, whose 
flux is tentatively shifted from 2.3 to 2.8$\times$10$^{-15}$\ergcs. 
The UV and mid-IR spectra are not fully specific to \IZ\,NW. 

An average de-reddened emission line spectrum for \IZ\,NW, close to 
the one secured by ICF99 in the optical range, is presented in Col.\,2 
(`Obs.') of Table~\ref{tab_res} (Line identifications 
in Col.\,1; Cols.\,3$-$6 are presented in Sect.~\ref{res}). 
This spectrum differs little from those by  Izotov \& Thuan (1998) 
and SK93. A rather deep, high-resolution red spectrum is presented 
by SK93. A few weak lines are taken from a deep blue MMT spectrum 
by TI05, who however quote an \oii\,3727 flux 
larger than in earlier studies. Absolute fluxes for \hb\ 
and the radio continuum are given on top of Table~\ref{tab_res}. 
The 21\,cm and 3.6\,cm fluxes, obtained from Cannon et al. (2005) 
as a sum of 3 contours for the NW shell, partly originate in non-thermal 
processes, not considered here. 
Line intensities are relative to \hb\,=\,1000. The intensity ratio 
\nii\la6584/\la6548 quoted by SK93 is smaller than the theoretical value: 
this is presumably due to the presence of a broad \ha\ component (VI98). 
Correcting for the pseudo-continuum, the theoretical ratio is 
recovered and a new, smaller value is obtained for the sum of the \nii\ 
doublet. \oii\,7320+30 (SK93) is uncertain and difficult to link to \ha. 
Taking into account weak (undetected) lines, such as \neiv\la4724, 
\feiii\la\la4702, 4734, 4755, a continuum slightly lower than the one 
adopted by TI05 leads to a 
moderate increase of the \ariv\ line fluxes.  Lines \feiv\,4906, 
\feii\,5158 and \fevi\,5176 are seen in the tracing by IT05, with 
tentative intensities 3, 2 and 2 (\hb\,=\,1000) respectively. Only \feiv\ 
is considered in Table~\ref{tab_res} (It is noted that the predicted 
intensities for these \feii\ and \fevi\ lines will be $\sim$\,1). 

The most critical (de-reddened) line ratio is $r$(\oiii) = 0.0246, the value 
also adopted by SS99. This is 3.1\% larger than the often quoted value by 
SK93 (2\arcsec slit), 0.6\% smaller than the value by ICF99 (1.5\arcsec slit) 
and 2.3\% smaller than in the blue spectrum by TI05 (2\arcsec slit).

\section{New photoionization models for \IZ\,NW}
\label{newmod}

Models are computed using the standard photoionization code {\sc nebu} 
(P\'equignot et al. 2001) in spherical symmetry with a central 
point-like source, suited to the apparent geometry of \IZ\,NW since the 
bulk of the stars of the NW MSC belongs to a cavity surrounded by the 
GEHIIR shell. Radiation-bounded filaments embedded in a diffuse medium 
are modelled. The reader is referred to Appendices~\ref{gas_distrib} 
\& \ref{WRstars} for a perspective to the present approach. Atomic data 
are considered in Appendix~\ref{fiat}. 

\subsection{Stellar ionizing radiation}
\label{mod_star}

The central source Spectral Energy Distribution (SED) is treated 
analytically, with no precise reference to existing synthetic stellar 
cluster SEDs (Appendix\,\ref{WRstars}). 
No effort is done to describe the optical+UV continuum. The continuum 
flux at \la3327\AA\ (de Mello et al. 1998) is not used to constrain the 
power of the MSC. Here, this constraint
can be replaced to great advantage by the fraction 
of ionizing photons absorbed in the shell (Sect.~\ref{abs_red}). 

The continuous distribution of stellar masses most often results in 
an approximately exponential decrease of flux with photon energy from 
1 to 4\,ryd in the SED of current synthetic MSCs (\eg, Luridiana et al. 2003). 
The sum of two black bodies at different temperatures can mimic this shape, 
yet providing flexibility to study the influence of the SED. 
The source of ionizing radiation is described as the sum of a hot 
black body, BB1 (temperature $T1$\,$\geq$\,60\,kK; luminosity $L1$), 
and a cooler one, BB2 ($T2$ = 40$-$50\,kK; $L2$). A constant 
scaling factor $\delta_4$ ($\leq$1), reminiscent of the discontinuity 
appearing in the SED of model stars (\eg, Leitherer et al. 1999) 
and constrained by the observed intensity of \heii\la4686, 
is applied to the BB1 flux at $\geq$~4\,ryd. 
The ionizing continuum depends on {\sl five free parameters}. 
The adopted $T2$ range is reminiscent of massive main sequence stars 
and lower $T2$'s need not be considered. A sufficiently large range of 
$T1$ values ought to be considered, as the high-energy tail of 
the intrinsic SED is influenced by quite a few WR stars, whose properties 
are either uncertain or unknowable (Appendix~\ref{WRstars}). 

\subsection{Ionized shell}
\label{mod_shell}

The \IZ\,NW shell extends from $R_{\rm i}$ = 2.85$\times10^{20}$\,cm to 
$R_{\rm f}$ = 4.75$\times10^{20}$\,cm (1.5\arcsec\ and 2.5\arcsec\ 
at $D$ = 4.0$\times$10$^{25}$\,cm). 

In `genuine' models a smooth small-scale density 
distribution is assumed (gas filling factor $\epsilon$ unity). 
The gas density is defined by means 
of the following general law for a variable gas pressure $P$, 
given as a function of the radial optical depth, $\tau$, at 13.6\,eV: 

\begin{equation}
P(\tau)=\frac{P_{\rm out}+P_{\rm in}}{2}+\frac{P_{\rm out}-P_{\rm in}}{\pi}
\tan^{-1}\Bigl[\kappa\log\Bigl(\frac{\tau}{\tau_{\rm c}}\Bigl)\Bigr].
\end{equation}

\noindent
This law is a convenient tool to explore the effects of the density 
distribution on the model predictions. 
$P$ is related to the pair of (\Te, \Nh) via the 
ideal gas law, with \Te\ derived from solving the statistical 
equilibrium equations at each step. At the first step of the computation 
($\tau = 0$), the initial pressure is $P_{\rm in}$, while at the last step 
($\tau = \tau_{\rm m} \sim \infty$) the final pressure is $P_{\rm out}$. 
A smooth, rapid transition is obtained here by adopting $\kappa = 30$ 
in all computations. Eq.\,1 introduces {\sl three free parameters}: 
$P_{\rm in}$, $P_{\rm out}$, and the optical depth $\tau_{\rm c}$ at 
which the transition from inner to outer pressure occurs. The picture 
of a filament core embedded in a dilute medium dictates that 
$P_{\rm in}$ $<$ $P_{\rm out}$. Each filament produces a radial shadow, 
which emits much less than the material in front of the filament and 
the filament itself, since it is only subject to the weak, very soft, 
diffuse field from the rest of the nebula. The shadows are neglected. 

In order to represent radiation-bounded filaments embedded 
in a low density medium (Appendix\,\ref{filling}), at least two 
sectors are needed: a `Sector~1' with $\tau_{\rm m}\gg\tau_{\rm c}$ 
(radial directions crossing a filament) and a `Sector~2' with 
$\tau_{\rm m} < \tau_{\rm c}$. To first order, only two sectors are 
considered. Observation shows that the \heii\ emission, although 
definitely extended, is relatively weaker in the filaments 
surrounding the main shell (VI98; 
Izotov et al. 2001). This deficit of \heii, unrelated to an 
outward decrease of the ionization parameter since \heii\ is a pure 
`photon counting' line above 4\,ryd, suggests instead that in no 
radial direction is the main shell totally deprived of absorbing gas. 
With the concern of reaching a more significant description, the same small 
$\tau_{\rm m}$(3) = 0.05 will be attached to the remaining `Sector~3' 
required to make up the covering factor of the source to unity in 
all complete models. The emission of Sector~3, a moderate contribution 
to the \heii\ intensity, does not impact on conclusions concerning 
the main shell and the source. 

For simplicity, in any given run, the values of the three defining 
parameters of Eq.\,1 are assumed to be shared by all three sectors. Note 
that $\tau_{\rm c}$ and $P_{\rm out}$ act only in Sector~1. The {\sl topology} 
(Appendix\,\ref{sphere_slit}) of the model shell is fully determined by 
specifying in addition the covering factors $f^{cov}_1$ of Sector~1 
(radiation-bounded) and $f^{cov}_2$ of Sector~2 (matter-bounded), 
with the condition: 

\begin{equation}
f^{cov}_3 = 1 - (f^{cov}_1 + f^{cov}_2) > 0, 
\end{equation}

\noindent
and finally the optical depth $\tau_{\rm m}$(2) ($< \tau_{\rm c}$) of 
Sector~2. The full model shell structure depends on {\sl six free parameters}. 

Adopting the same $R_{\rm i}$ and the same parameters for 
$P(\tau)$ in the three sectors and assuming that the outer radius of 
Sector~1 is $R_{\rm f}$ make the computed outer radii of other sectors 
to be smaller than $R_{\rm f}$. If, however, one would like Sector~2 
to extend up to $R_{\rm f}$ and perhaps beyond, models should be re-run 
for this sector using now a $P(\tau)$ with $P_{\rm out}(2) < P_{\rm in}$ 
and $\tau_{\rm c}$\,$<$\,1. No significant 
consequences on the computed spectrum result from this change, as the 
increase of radius and the decrease of density in the outermost layers 
of Sector~2 (say, $\tau$\,$\sim$\,1) have opposite effects on the `local' 
ionization. Also, `improving' the artificial geometry of Sector~3 
(a thin shell at radius $R_{\rm i}$) by assuming a lower $P_{\rm in}$(3) 
or else a filling $\epsilon\ll 1$ would not change at all the intensity 
of \heii, while the emission of other lines from this sector is negligible. 

Although three sectors are considered, 
Sector~3 is of no practical consequence for the main shell and no 
parameter is attached to it. A model based on the above description 
will be termed a `two-sector model' (Sect.~\ref{res_M2comp}).

\begin{table}
%Table 1
\caption[]{Constraints on model parameters$^a$}
\label{tab_conv}
%\centering                          % used for centering table
\begin{tabular}{l|l}
\hline\hline
Parameter  &  Constraints                \\
[0.1cm]
\hline                        % inserts single horizontal line
  %% \cline{1-2} 
E(B-V)     & \ha/\hb; (No freedom: E(B-V) = 0.04) \\
$R_{\rm i}$ & No freedom $R_{\rm i}$ = 2.85$\times10^{20}$\,cm \\
$f^{cov}$  & Absolute I(\hb) = 5.6$\times$10$^{-14}$\ergcs \\
$L1$, $L2$ & Cluster SED; $f^{cov} \leq 1.0$; $Q_{\rm abs}$/$Q$\,$\sim$\,0.37\\
$T1$, $T2$ & $\delta_4 \leq 1.0$; (log($Q_{\rm He}$/$Q$)\,$\sim$\,$-$0.5) \\
$\delta_4$ & \heii\la4686 \\
[0.1cm]
\cline{1-2}
He & He/H = 0.08; \hei\la5876? \\
C  & \ciii\la1909  \\
N  & \nii\la6584   \\
O  & \oiii\la5007  \\
Ne & \neiii\la3869 \\
Mg &  Mg/Ar = 10.; \mgi\la4571?  \\
Al &  Al/Ar = 1.; \aliii\la1855? \\
Si &  Si/Ar = 10.; \sit\la1883?  \\  
S  &  S/Ar = 4.37; $<$\siii$>$? \\
Ar & \ariii\la7135 \\ 
Fe & \feiii\la4658 \\
[0.1cm]
\cline{1-2} 
\multicolumn{2}{l}{One-component constant density run ($N0$, $N1$):} \\
\Nh            & $r$(\sii) = \sii\la6716/\sii\la6731 ($\pm$) \\
$\tau_{\rm m}$ & \oii\la3727 \\
$\epsilon$     & $R_{\rm f}$ = 4.75$\times10^{20}$\,cm \\
[0.1cm]
\cline{1-2} 
\multicolumn{2}{l}{One-component model with Eq. (1) ($M1$):} \\
$\epsilon = 1.0$ & No freedom \\
$P_{\rm out}$  & $r$(\sii) \\
$\tau_{\rm c}$ & \oii\la3727 \\
$P_{\rm in}$   & $R_{\rm f}$ = 4.75$\times10^{20}$\,cm \\
[0.1cm]
\cline{1-2} 
\multicolumn{2}{l}{Two-component model ($M2, M3, M4$) added freedoms:} \\
$f^{cov}_1$ & $f^{cov}_2$ $>$ 0; $f^{cov}$ = $f^{cov}_1$ + $f^{cov}_2$ $<$ 1 \\
$f^{cov}_2$ & $\tau_{\rm m}$(2) $<$ $\tau_{\rm c}$; global 3-D geometry \\
$\tau_{\rm m}$(2) & Fine tuning I(\hb) for given $f^{cov}_i$\\
[0.1cm]
\hline
\end{tabular}

\ $^a$Question marks attached to dismissed constraints (see text).\\
\end{table}

\subsection{Model parameters and constraints}
\label{mod_constraints}

Correspondances between model parameters and constraints are outlined in 
Table~\ref{tab_conv}. The parameters are interrelated and iterations are 
needed to converge to a solution. The weak dependance of E(B-V) on the model 
Balmer decrement (Sect.~\ref{abs_red}) is neglected. The SED 
is not fully determined by the major constraint $Q_{\rm abs}$/$Q$. Other 
constraints are in the form of inequalities, some are half-quantitative 
or deal with `plausibility' arguments. 

One emission line is selected to constrain each elemental abundance. 
In Table~\ref{tab_conv}, 
a question mark is appended to those lines with unreliable intensities
(Table~\ref{tab_res}): the intensity of \mgi\la4571+62 is given as 
an upper limit as the lines are barely detected (TI05) and suspected 
to be blended with a WR feature (Guseva et al. 2000); detection of 
\aliii\la1855 is a simple guess from a tracing of the {\sc hst} spectrum; 
\sit\la1883+92 is barely seen in the {\sc iue} spectrum and only the 
first component of the doublet is detected in the {\sc hst} spectrum 
(IT99). The abundances of Mg, Al and Si are arbitrarily linked to that of 
argon (Table~\ref{tab_conv}), assuming abundance ratios close to solar 
(Lodders 2003). For simplicity, the solar S/Ar ratio is also adopted and 
the computed sulfur line intensities can be used to scale S/H according to 
any preferred criterion (Sect.~\ref{disc_ab}). 
\hei\ emission lines are blended with strong stellar absorption lines 
(\eg, ICF99). He/H is set at 0.08 by number. 

\begin{table}
%\begin{table*}
%Table 2
\caption[]{\IZ\,NW: model parameters and properties}
\begin{tabular}{l|c|rrrr}
\hline\hline
Parameters            & Run$^a$ & \multicolumn{4}{c}{Model$^b$} \\
of model              & N0-N1    & M1    & M2   &  M3   & M4   \\
                      & 2 $-$ 3  &  4    &  5   &   6   &  7   \\
\cline{1-6}
\multicolumn{6}{l}{\underline{Central source parameters}}      \\
$T1$/10$^4$\,K        &  10.     &  10.  &  8.   &  8.   &  12. \\
$L1$/10$^{41}$\ergs   & 3.5-1.6  & 3.5   & 2.0   & 1.6   & 1.25 \\
$\delta_4$            & .12-.67  & 0.73  & 0.83  & 0.93  &0.24 \\
$T2$/10$^4$\,K        &   4.     & 4.    &  4.   &   5.  &  4. \\
$L2$/10$^{41}$\ergs   & 3.5-1.6  &  3.5  & 2.0   &  1.6  &  2.5 \\
log($Q$) $-$ 51.      & 1.04-.71 & 1.054 & 0.829 & 0.779 & 0.734\\ %1.045
$-$log($Q_{\rm He}$/$Q$) &.462-.447 &0.445 &0.523 &0.493 &0.539 \\ 
\multicolumn{6}{l}{\underline{Ionized shell parameters}}        \\
$\epsilon$            &  .0042-.31  & 1.00 & 1.00  & 1.00 & 1.00 \\
$P_{\rm in}/k$/10$^5${\sc cgs} &38.-8.3 & 5.01 & 3.40 & 2.96 & 2.71 \\
$P_{\rm out}/k$/10$^5${\sc cgs} & - & 23.2 & 21.7 & 25.4 & 26.8 \\
$\tau_{\rm c}$         &     -  & 5.7  &  4.9 & 4.0  & 4.3  \\
$f^{cov}_1$           & 1.-0.43 & 0.20 & 0.26 & 0.23 & 0.29 \\
$f^{cov}_2$           &    -    & -    & 0.30 & 0.50 & 0.60 \\ 
$\tau_{\rm m}$ or $\tau_{\rm m}$(2)     
                      & .96-300.& 270. & 1.21 & 1.46 & 1.00 \\ 
\multicolumn{6}{l}{\underline{Elemental abundances by number (H = 10$^7$)}} \\
C                     &  60-52  & 45.8 & 38.8 & 35.3 & 47.4 \\ 
N                     & 8.1-5.3 &  3.9 &  4.1 &  4.0  & 3.8 \\
O                     & 198-192 & 172. & 168. & 162. & 173. \\ 
Ne                    & 33-30   & 26.4 & 25.7 & 24.8 & 26.9 \\
S (Table~\ref{tab_conv})& 3.4-3.9 &  5.0 & 4.3 & 4.3 &  5.0 \\
Ar                    & .77-.90 & 1.14 & 0.99 &  0.98& 1.13 \\
Fe                    & 4.8-5.8 &  5.8 &  6.1 &  6.0 &  6.5 \\
\multicolumn{6}{l}{\underline{Mean shell properties weighted by \Ne$\times$$N_{\rm ion}$, except for \Ne}}\\
$Q_{\rm abs}$/$Q$     & .20-.43   & .200  & .343 & .380 & .426 \\ %.202
$M_{\rm gas}$/10$^6$\Msun& .15-.92 & 1.02 & 1.52 & 1.74 & 1.79 \\
H$^{+}$/H             & .998-.98  & .957  & .961 & .963 & .948 \\
O$^{0}$/O             & .00-.017  & .041  & .038 & .037 & .052 \\
O$^{+}$/O             & .076-.10  & .122  & .125 & .129 & .131 \\
O$^{2+}$/O            & .910-.85  & .784  & .793 & .790 & .773 \\
O$^{3+}$/O            & .014-.03  & .049  & .043 & .043 & .042 \\
\Te(H$^{+}$)/10$^4$K  & 1.65-1.73 & 1.873 &1.859 &1.896 &1.839 \\
\Te(O$^{0}$)/10$^4$K  & 1.61-1.14 & 1.040 &1.013 &1.012 &1.003 \\
\Te(O$^{+}$)/10$^4$K  & 1.63-1.43 & 1.320 &1.315 &1.309 &1.272 \\ 
\Te(O$^{2+}$)/10$^4$K & 1.66-1.75 & 1.911 &1.915 &1.961 &1.898 \\
\Te(O$^{3+}$)/10$^4$K & 1.76-2.1  & 2.576 &2.402 &2.449 &2.479 \\
\Ne(H$^{+}$)/\cc      &    99-18  & 17.3  & 11.1 &  9.8 &  9.4 \\
\Ne(O$^{0}$)/\cc      &   99-8.5  & 46.7  & 34.0 & 37.8 & 41.1 \\
\Ne(O$^{+}$)/\cc      &    99-16  & 62.7  & 48.0 & 54.0 & 54.8 \\
\Ne(O$^{2+}$)/\cc     &    99-18  & 15.4  & 10.2 &  8.9 &  8.5 \\
\Ne(O$^{3+}$)/\cc     &   100-19  & 10.1  & 7.4  &  6.3 &  5.7 \\
$t^2$(H$^{+}$)        & .002-.014 & .032 & .026 & .028 & .030 \\ %.0018
$t^2$(O$^{0}$)        & .001-.018 & .022 & .024 & .023 & .020 \\ %.0009
$t^2$(O$^{+}$)        & .0013-.02 & .021 & .024 & .024 & .023 \\ %.0013
$t^2$(O$^{2+}$)       & .002-.008 & .013 & .010 & .010 & .010 \\ %.0018
$t^2$(O$^{3+}$)       & .0010-05  &.0008 &.0006 &.0007 &.0016 \\
[0.1cm]
\hline
\end{tabular}

\ \ $^a$Constant \Nh. $N0$: \Nh\,=\,92\cc; $N1$: \Nh\,=\,17.0\cc.\\
\ \ $^b$\Nh\ from thermal pressure of Eq. (1).\\
%% S$^c$ \ \ $^c$Mg, Al, Si, S: see Table~\ref{tab_conv}. \\
 
\label{tab_mod}
\end{table}

In the lower part of Table~\ref{tab_conv} are given observational 
constraints for the structural parameters of the shell, depending on 
assumptions. In preliminary constant-density `runs' ($N0$, $N1$, 
not genuine models; Sect.~\ref{res_N}), a generalization of the approach 
of SS99 (Sect.~\ref{prev_recent}) is adopted. A one-component model 
($M1$; Sect.~\ref{res_M1comp}) shows the influence of Eq. (1). 
Two-component models ($M2$, $M3$ and $M4$; Sect.~\ref{res_M2comp}) 
generalize $M1$ according to Sect.~\ref{mod_shell}.

\section{Results}
\label{res}

\begin{figure}
%Fig.1
\resizebox{\hsize}{!}{\includegraphics{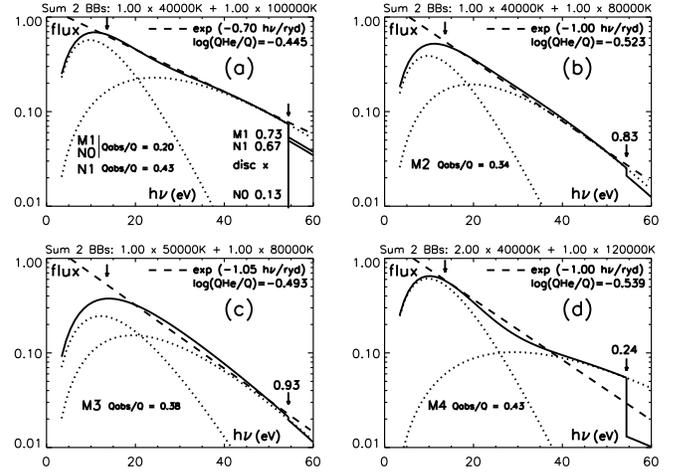}}
\caption[]{
Spectral energy distribution (SED) for \IZ\,NW models. Flux in \ergs\,eV$^{-1}$ multiplied by an arbitrary constant {\sl versus} photon energy $h\nu$ in eV. The SED (solid line) is the sum of two black bodies (dotted lines), with a discontinuity at 4 ryd (Sect.~\ref{mod_star}). Panel (a) is common to $N0$, $N1$ and $M1$. Panels (b), (c) and (d) correspond to $M2$, $M3$ and $M4$ respectively. Vertical arrows appear at 1 and 4 ryd. Straight dashed lines of the form exp($-\alpha$\,$h\nu$/ryd) are exponential approximations to the SEDs over the range 1$-$4 ryd to help eyes. The exponential expressions are written in the panels corresponding to the SEDs ($\alpha$ = 0.70, 1.00, 1.05 and 1.00 in panels (a) to (d), with $\alpha$ a measure of the EUV continuum hardness). Some of the SED data provided in Table~\ref{tab_mod} ($Q_{\rm He}$/$Q$, discontinuity factor $\delta_4$ and $Q_{\rm abs}$/$Q$) are recalled in the panels.
}
\label{fig_SED}
\end{figure}

\begin{figure}
%Fig.2
\resizebox{\hsize}{!}{\includegraphics{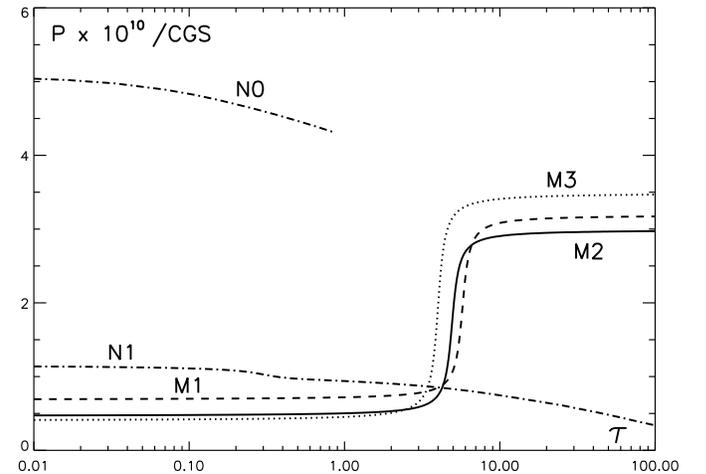}}
\caption[]{Gas pressure {\sl versus} optical depth $\tau$ for models $M1$ (dashed line), $M2$ (solid line) and $M3$ (dotted line), governed by Eq.~1 with parameters in Table~\ref{tab_mod}. $M4$, close to $M3$, is not shown. For comparison, the dash-dotted lines correspond to the constant density runs $N0$ and $N1$. 
}
\label{fig_P}
\end{figure}

\begin{table}
%\begin{table*}
%Table 3
\caption[]{Model outputs for \IZ\,NW: $N0$ and $M2$}
\begin{tabular}{l|r|rc|rc}
\hline\hline
Line id./Models & Obs.  & N0 & N0/O  &  M2  & M2/O\\
\cline{1-6}
\multicolumn{5}{l}{\ \underline{Absolute fluxes} (I(\hb) in 10$^{-14}$\ergcs)}\\
I(\hb)           & 5.6   & 5.6  & 1.00 &  5.6  & 1.00 \\
1.43$\:$GHz$\:$/mJy  & 0.433 & 0.240& 0.55 & 0.256 & 0.59 \\
8.45$\:$GHz$\:$/mJy  & 0.286 & 0.202& 0.71 & 0.216 & 0.76 \\
\multicolumn{5}{l}{\ \underline{Relative line fluxes} (wavelengths in \AA\ or $\mu$m)} \\
 \hi\ 4861       & 1000. &1000. & 1.00 & 1000. & 1.00 \\
 \hi\ 6563       & 2860. &2840. & 0.99 & 2830. & 0.99 \\
 \hi\ 4340       &  461. & 473. & 1.03 & 473.  & 1.03 \\
 \hi\ 4102       &  266. & 267. & 1.00 & 267.  & 1.00 \\
 \hi\ 1215 (/10)   &  -  &2950. &  -   & 3010. &  -   \\
 \hi\ 2$h\nu$ (/10)&  -  &1550. &  -   & 1620. &  -   \\
 \hei\ 3888      &  90.4 & 89.6 & 0.99 &  90.5 & 1.00 \\
 \hei\ 4471      &  21.4 & 34.9 & 1.63 &  35.2 & 1.64 \\
 \hei\ 5876      &  67.7 & 92.0 & 1.36 &  91.3 & 1.35 \\
 \hei\ 6678      &  25.3 & 25.3 & 1.00 &  25.6 & 1.01 \\
 \hei\ 7065      &  24.4 & 23.4 & 0.96 &  22.9 & 0.94 \\
 \hei\ 10830     &   -   & 251. & -    & 190.  & -    \\
 \heii\ 4686     &  36.8 & 36.8 & 1.00 &  36.8 & 1.00 \\
 {\ciii} 1909+07 & 467.  & 467. & 1.00 & 467.  & 1.00 \\
 {\sit} 1882+92  & 270.: & 229. & 0.85 & 340.  & 1.26 \\
 {\aliii} 1855+63 & 111.:& 42.9 & 0.39 &  82.9 & 0.75 \\
 {\oiiis} 1664    &$<$230 & 127. &  $>$.5 & 208. & $>$.9\\
 {\civ} 1549      & 510.:& 74.2 & 0.14 & 334.  & 0.65 \\
 {\siq} 1397 \ $\rbrace$$^a$ &$<$300&22.7& $>$.1 &127. & $>$.5\\
 {\oivs} 1398  $\rbrace$ & * & 2.0 &  *   & 24.9 &  *   \\
 {\nii} 6584+48      &  9.2 &  9.2 & 1.00 &  9.2 & 1.00 \\
 {\oi} 6300+63       &  8.5 & 0.12 & 0.01 &  8.6 & 1.01 \\
 {\oii} 3726+29      & 238. & 238. & 1.00 & 238. & 1.00 \\
 {\oii} 7320+30      & 6.3: & 9.5  & 1.50 &  7.5 & 1.18 \\
 {\oiii} 5007+..     &2683. &2680. & 1.00 &2680. & 1.00 \\
 {\oiii} 4363        & 65.9 & 47.8 & 0.73 & 63.2 & 0.96 \\
 {\oiii} 51.8$\:\mu$m &  -  & 174. &   -  & 137. &  -   \\
 {\oiii} 88.3$\:\mu$m &  -  & 216. &   -  & 213. &  -   \\
 {\oiv} 25.9$\:\mu$m  &49.1 & 18.2 & 0.37 & 47.8 & 0.97 \\
 {\neii} 12.8$\:\mu$m &9.0: & 1.3  & 0.14 &  1.9 & 0.21 \\
 {\neiii} 3868+..    & 191. & 191. & 1.00 & 191. & 1.00 \\
 {\neiii} 15.5$\:\mu$m& 45.7& 60.0 & 1.31 & 48.9 & 1.07 \\
 {\mgi} 4571+62     &$<$3.0 &  1.5 &$>$.5 &  1.2 & $>$.4 \\
 {\sid} 34.8$\:\mu$m & 157. &  4.7 & 0.03 & 22.0 & 0.14 \\
 {\sii} 6716         & 22.5 &  6.7 & 0.30 & 17.6 & 0.78 \\
 {\sii} 6731         & 16.9 &  5.1 & 0.30 & 13.1 & 0.78 \\
 {\sii} 4068 \ $\:\rbrace$& 3.7& 1.1& 0.41&  2.2 & 0.99 \\
 {\fev} 4071 $\rbrace$ &  * & 0.4  &   *  &  1.5 &  *   \\
 {\siii} 9531+..     & 114. & 130. & 1.15 & 113. & 0.99 \\
 {\siii} 6312        &  6.7 &  6.0 & 0.89 &  5.7 & 0.85 \\
 {\siii} 18.7$\:\mu$m& 28.0 & 32.2 & 1.15 & 26.2 & 0.95 \\
 {\siii} 33.5$\:\mu$m& 120. & 54.5 & 0.45 & 48.0 & 0.40 \\
 {\siv} 10.5$\:\mu$m & 48.0 & 41.7 & 0.87 & 92.6 & 1.93 \\
 {\ariii} 7136+..    & 23.5 & 23.5 & 1.00 & 23.5 & 1.00 \\
 {\ariii} 8.99$\:\mu$m &  - &  8.6 &   -  &  8.1 &  -   \\
 {\ariv} 4711 $\rbrace$& 8.6 & 1.5 & 0.76 &  8.2 & 1.53 \\
 {\hei} 4713 \ \ $\:\:\rbrace$& *&5.0&  * &  5.0 &  *   \\
 {\ariv} 4740          & 4.5 & 1.2 & 0.26 &  6.2 & 1.39 \\
 {\feii} 5.34$\:\mu$m  &  -  & 0.1 & -    & 10.8 &   -  \\
 {\feii} 26.0$\:\mu$m  & 34. & 0.0 & 0.00 &  3.4 & 0.10 \\
 {\feiii} 4658         & 4.5 & 4.5 & 1.00 &  4.5 & 1.00 \\
 {\feiii} 4986         & 7.4 & 5.8 & 0.78 &  7.0 & 0.94 \\
 {\feiii} 22.9$\:\mu$m &  -  & 2.2 &   -  &  3.2 &   -  \\
 {\feiv} 4906         & 3.0: & 1.6 & 0.54 &  3.1 & 1.02 \\
 {\fev} 4227           & 1.8 & 1.5 & 0.84 &  5.5 & 3.10 \\
\hline
\end{tabular}

\ \ $^a$ In Col.\,1, blends are indicated by braces. The observed intensity of 
a blend is attributed to the first line and an asterisk to the second line.\\

\label{tab_res}
\end{table}

\begin{table}
%Table 4
\caption[]{Model `predictions' for \IZ\,NW: model intensities divided by observed intensities}
\begin{tabular}{l|c|c|cccc}
\hline\hline
 Line ident.           & Obs. &  N0$-$N1  &  M1  &  M2  &  M3  &  M4  \\
                       &  2   &  3 $-$ 4  &  5   &  6   &  7   &  8   \\
\cline{1-7}
 $Q_{\rm abs}$/$Q$$^a$ & 0.37 & .20$-$.43 & 0.20 & 0.34 & 0.38 & 0.43 \\
\cline{1-7}
 {\civ}\ 1549          & 500: & .14$-$.41 & 1.12 & 0.65 & 0.64 & 1.08 \\
 {\oi}$\:$6300+        &  8.5 & .01$-$.59 & 1.19 & 1.01 & 0.95 & 1.38 \\
 {\oii}$\:$7320+      & 6.3: & 1.5$-$1.24 & 1.19 & 1.18 & 1.19 & 1.15 \\
 {\oiii}$\:$4363       & 65.9 & .73$-$.82 & 0.96 & 0.96 & 1.00 & 0.94 \\
 {\oiv}$\:$25.9$\mu$   & 49.1 & .37$-$.84 & 1.08 & 0.97 & 0.94 & 0.98 \\
 {\neii}$\:$12.8$\mu$  & 9.0: & .14$-$.10 & 0.14 & 0.21 & 0.22 & 0.23 \\
 {\neiii}15.5$\mu$   & 45.7 & 1.31$-$1.22 & 1.09 & 1.07 & 1.02 & 1.11 \\
 {\sid}$\:$34.8$\mu$   & 157. & .03$-$.10 & 0.17 & 0.14 & 0.14 & 0.18 \\
 {\sit}$\:$1882+      & 270: & .85$-$.99 & 1.12 & 1.26 & 1.34 & 1.16 \\
 {\sii}$\:$6716        & 22.5 & .30$-$.54 & 0.96 & 0.78 & 0.77 & 1.10 \\
 {\sii}$\:$6731        & 16.9 & .30$-$.51 & 0.96 & 0.78 & 0.77 & 1.10 \\
 {\sii}$\:$4068$^b$ & 3.7$^b$ & .30$-$.43 & 0.73 & 0.58 & 0.58 & 0.80 \\
 {\siii}$\:$9531+    & 114. & 1.15$-$1.05 & 0.95 & 0.99 & 1.00 & 0.92 \\
 {\siii}$\:$6312       &  6.7 & .89$-$.84 & 0.77 & 0.85 & 0.87 & 0.75 \\
 {\siii}$\:$18.7$\mu$ & 28. & 1.15$-$1.01 & 0.94 & 0.95 & 0.94 & 0.91 \\
 {\siii}$\:$33.5$\mu$  & 120. & .45$-$.44 & 0.39 & 0.40 & 0.40 & 0.39 \\
 {\siv}$\:$10.5$\mu$   & 48. & .87$-$1.50 & 2.15 & 1.93 & 1.97 & 2.15 \\
 {\ariv}$\:$4740       &  4.5 & .26$-$.73 & 1.87 & 1.39 & 1.47 & 1.76 \\
 {\feii}$\:$26.0$\mu$  &  34. & .00$-$.05 & 0.10 & 0.10 & 0.10 & 0.14 \\
 {\feiii}$\:$4986      & 7.4 & .78$-$1.15 & 0.90 & 0.94 & 0.89 & 0.89 \\
 {\feiv}$\:$4906       & 3.0: & .54$-$.70 & 0.96 & 1.02 & 1.09 & 1.02 \\
 {\fev}$\:$4227        & 1.8: & .84$-$2.3 &  2.1 &  3.1 &  3.2 &  3.1 \\
[0.1cm]
\hline
\end{tabular}

\ \ $^a$$Q_{\rm abs}$/$Q$: absolute values.\\
\ \ $^b$\la4068 intensity not corrected for \fev\la4071 (Table~\ref{tab_res}). 
 
\label{tab_compa}
\end{table}

Input and output model properties are listed 
in the first column of Table~\ref{tab_mod} as: 
(1) five primary ionizing source parameters (Sect.~\ref{mod_star}); 
(2) resulting numbers of photons (s$^{-1}$) $Q$ and $Q_{\rm He}$ 
emitted by the source above 13.6 and 24.6\,eV respectively; 
(3) four ($N0$, $N1$, $M1$) to six ($M2$, $M3$, $M4$) shell 
parameters (Sect.~\ref{mod_shell});  
(4) elemental abundances; 
(5) photon fraction $Q_{\rm abs}$/$Q$ absorbed in the shell; 
(6) mass $M_{\rm gas}$ of ionized gas in units of 10$^6$\Msun; 
(7) mean ionic fractions of H$^+$ and oxygen ions weighted by \Ne;  
(8) average \Te\ and $t^2$ weighted by \Ne$\times$$N_{\rm ion}$ and 
average \Ne\ weighted by $N_{\rm ion}$ for H$^+$ and oxygen ions. 

The model SEDs (Sect.~\ref{mod_star}) are shown in Fig.~\ref{fig_SED}:  
panel (a) is common to $N0$, $N1$ and $M1$; panels (b), (c) and (d) 
correspond to $M2$, $M3$ and $M4$ respectively. Radiation is harder and 
stronger in panel (a) (see `hardness coefficient' 
$\alpha$ in caption to Fig.~\ref{fig_SED}). 
The gas pressure laws $P(\tau)$, drawn in Fig.~\ref{fig_P} 
(parameters in Table~\ref{tab_mod}), illustrate the 
contrast between preliminary runs and adopted models. 

Line identifications and observed de-reddened intensities are  
provided in  Cols.\,1 and 2 of Table~\ref{tab_res}. 
Computed intensities appear in Col.\,3 and Col.\,5 for Run $N0$ 
and Model $M2$ respectively. Predictions are given for some unobserved 
lines (intensities are 10 times the quoted values for \hi\,1215\AA\ 
and 2$h\nu$). The ratios of computed to observed intensities, 
noted `$N0$/O' and `$M2$/O' appear in Col.\,4 and Col.\,6 for $N0$ 
and $M2$ respectively. Ideally, these ratios should be 1.00 for all 
observed lines. 

Inasmuch as the convergence is completed, at least {\sl all} lines which 
were used as model constraints (Table~\ref{tab_conv}) must be exactly 
matched by construction (Table~\ref{tab_res}). For the 
sake of evaluating the models, these lines are therefore useless. 
Similarly, `redundent' lines (\hi\ and \hei\ series, etc.), which carry 
no astrophysically significant information in the context, as well as 
unobserved lines, can be discarded. Remaining `useful' lines are 
listed in Cols.\,1$-$2 of Table~\ref{tab_compa} and model intensities 
divided by observed intensities are displayed in Cols.\,3$-$8 for $N0$--$N1$, 
$M1$--$M4$ respectively. These intensities are `predictions' in that they 
are not considered at any step of the convergence. 
In  Table~\ref{tab_compa}\footnote{Since 
\oiii\la5007+4959 is exactly matched, 
the entry \oiii\la4363 in Tables~\ref{tab_compa} and \ref{tab_vari} 
is the `normalized $r$(\oiii)', \ie, the ratio of the computed 
$r$(\oiii) to the observed $r$(\oiii).}, \oiii\la4363\AA\ stands out as 
the strongest, accurately measured optical line. 
$Q_{\rm abs}$/$Q$ is repeated in Table~\ref{tab_compa}. 

\subsection{Constant density runs with filling factor: $N0$, $N1$}
\label{res_N}

$N0$ is a preliminary run (Col.\,2 of Table~\ref{tab_mod}) 
in which \Nh\ is constant and $Q$, $Q_{\rm He}$, 
$R_{\rm i}$ and $R_{\rm f}$ are about as in the description 
by SS99 (corrected for the larger $D$). The convergence 
process, involving O/H, Ne/H, etc. (Table~\ref{tab_conv}), 
is more complete than the one performed by SS99, but the 
differences of procedures do not change the conclusions. 
If \Nh\ is in principle derived from $r$(\sii), the 
sensitivity of $r$(\sii) to \Ne\ is relatively weak at the low 
density prevailing in the shell, while the exact value adopted 
for \Nh\ may, {\sl in this particular structure}, strongly 
influence the computed spectrum. By changing coherently \Nh, 
$\epsilon$ and $f^{cov}$, the three constraints I(\hb), 
\oiii/\oii\ and $R_{\rm f}$ can be fulfilled along a sequence. 
$N0$ is extracted from this sequence by assuming, as in the 
SS99 run, a covering factor $f^{cov}$ = 1. The solution is close to 
the one chosen by SS99, with \Nh\,=\,92\cc, $\epsilon$\,=\,0.0042, 
(radial) $\tau$\,$\sim$\,1 ($\equiv$ $\tau_{\rm m}$(2) in 
Table~\ref{tab_mod}) and $r$(\sii) only $-$2.1\% off the observed value. 
$N0$ (Cols.\,3$-$4 of Table~\ref{tab_res}; Col.\,3 of Table~\ref{tab_compa}) 
{\sl fully confirms the very large problems met by SS99} 
with \oiii\la4363 and \oi\la6300 (Sect.~\ref{prev_recent}). 

\begin{figure}
%Fig.3
\resizebox{\hsize}{!}{\includegraphics{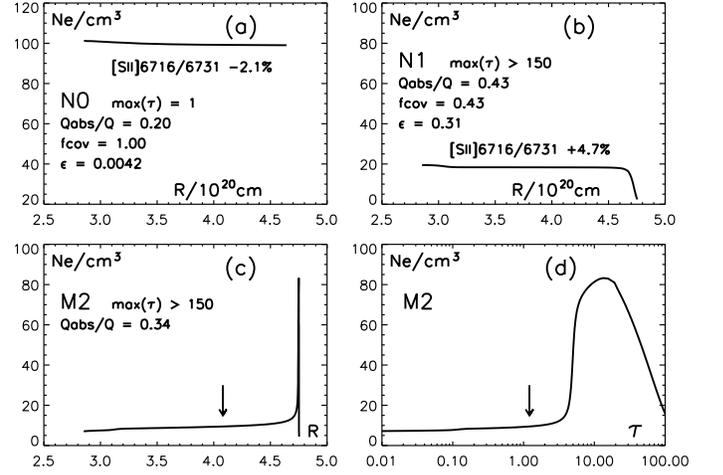}}
\caption[]{
Local electron density \Ne\ {\sl versus} shell radius $R$ in \IZ\,NW models: (a) $N0$, (b) $N1$, (c) $M2$; and {\sl versus} optical depth $\tau$ at 1~ryd: (d) $M2$. In panels (c) and (d), the vertical arrows indicate the boundary of the matter-bounded sector 2 of Model $M2$. Also provided are the optical thickness `max($\tau$)' (`$>$ 150' means `radiation bounded'), the fraction of primary photons absorbed $Q_{\rm abs}$/$Q$, and, where relevant ($N0$ and $N1$), the covering factor $f^{cov}$, the filling factor $\epsilon$ and the departure of the computed \sii\ doublet ratio from observation in \%. 
}        
\label{fig_Ne}
\end{figure}

\begin{figure}
%Fig.4
\resizebox{\hsize}{!}{\includegraphics{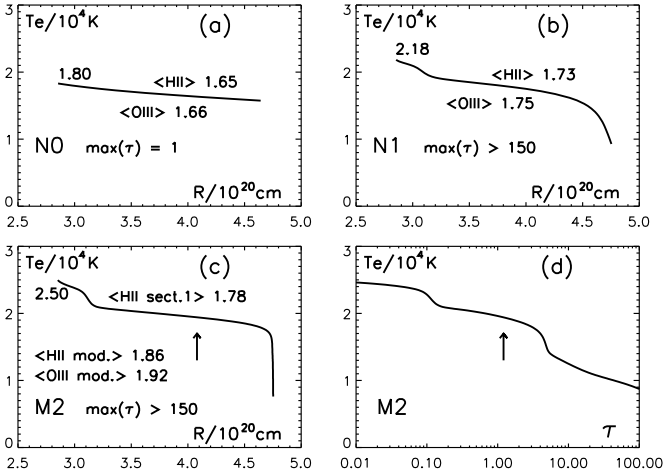}}
\caption[]{
As in Fig.~\ref{fig_Ne} for local electron temperature \Te/10$^4$\,K. The maximum \Te\ and average \Te's weighted by \Ne$\times$N(H$^+$) and by \Ne$\times$N(O$^{2+}$) are noted along each curve. In panel ($c$), $<$\Te$>$ is given for both the radiation bounded sector ($<$sect.1$>$) and the full model $M2$ ($<$mod.$>$).}
\label{fig_Te}
\end{figure}

\begin{figure}
%Fig.5
\resizebox{\hsize}{!}{\includegraphics{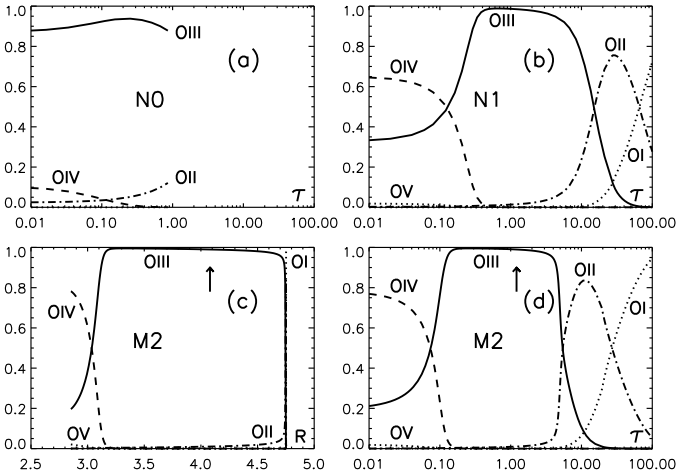}}
\caption[]{
Local fractional concentrations of O$^{0}$ (dotted line), O$^{+}$ (dot-dashed line), O$^{2+}$ (solid line), O$^{3+}$ (dashed line) and O$^{4+}$ (dotted line again) {\sl versus} $\tau$ in \IZ\,NW models: (a) $N0$, (b) $N1$, (d) $M2$; and {\sl versus} $R$: (c) $M2$. Vertical arrows as in Fig.~\ref{fig_Ne}. 
}
\label{fig_Oxy}
\end{figure}

$N0$ also fails in that $Q_{\rm abs}$/$Q$ is half the expected value. 
Decreasing $Q$ (SED luminosity) implies to decrease \Nh\ (for \oiii/\oii) 
and increase $\epsilon$ (for I(\hb)). Decreasing \Nh\ should help 
increasing \Te, thus $r$(\oiii), and the high ionization lines, 
largely underestimated in $N0$. Correlatively, \oiii/\oii\ is restored 
for a larger $\tau_{\rm m}$, which helps increasing \oi\ and other 
low-ionization lines. By further decreasing \Nh\ and increasing $\epsilon$, 
whilst fine-tuning $f^{cov}$ to keep the outer shell radius $R_{\rm f}$ 
(and I(\hb)) and $\delta_4$ (for \heii), it is possible to further 
increase $\tau_{\rm m}$ until the shell 
eventually gets {\sl radiation bounded}. The resulting (unique) solution 
is $N1$ (Col.\,3 of Table~\ref{tab_mod}; Col.\,4 of Table~\ref{tab_compa}), 
which much improves upon $N0$ concerning \oi\ and high-ionization lines, while 
$Q_{\rm abs}$/$Q$ $\equiv$ $f^{cov}$ = 0.43 is just a bit large. Because 
of the much lower density, \Nh\,=\,17\cc, the ratio $r$(\sii) is now $+$4.7\% 
off, worse than in $N0$, yet not decisively inacceptable. Nevertheless, the 
normalized $r$(\oiii), enhanced from 0.73 to 0.82, is still very significantly 
{\sl too small}. This failure of $N1$ is illustrated in the upper panels 
of Figs.~\ref{fig_Ne}$-$\ref{fig_Oxy}. The runs of \Ne\ and \Te\ with 
nebular radius $R$ are shown in Figs.~\ref{fig_Ne}a--\ref{fig_Te}a for 
$N0$ and Figs.~\ref{fig_Ne}b--\ref{fig_Te}b for $N1$. 
Ionic fractions of oxygen {\sl versus} $\tau$ are shown in 
Figs.~\ref{fig_Oxy}a and \ref{fig_Oxy}b for $N0$ and $N1$ respectively. 
In $N0$, \Te\ is everywhere above 1.55$\times 10^{4}$\,K. In $N1$, 
the inner \Te\ is 3\,800\,K higher than in $N0$, but O$^{2+}$ is 
abundant up to $\tau$ = 15, where \Te\ is below 1.40$\times 10^{4}$\,K 
and the {\sl average} \Te(\oiii) is not much increased. 

This generalization shows that no solution with 
constant \Nh\ exists even for the rather hard 
SED adopted by SS99. In an extreme variant of $N1$, the SED is just one 
10$^5$\,K black body (converged $L$ = 2.7$\times$10$^{41}$\ergs, 
$\delta_4$ = 0.41, log($Q_{\rm He}$/$Q$) = $-$0.27), but the normalized 
$r$(\oiii) = 0.87 is still too small, despite the unrealistically hard SED. 

\subsection{A one-sector photoionization model: $M1$}  
\label{res_M1comp}

Model $M1$ includes the same primary source as Run $N0$ and 
again only one sector, but with \Nh\ controlled by Eq.~1 (Fig.~\ref{fig_P}). 
Parameters appear in Col.\,4 of Table~\ref{tab_mod} and predictions in Col.\,5 
of Table~\ref{tab_compa}. The litigious lines \oiii\la4363 and \oi\la6300 
are {\sl very much improved} compared to Run $N0$ (and even $N1$), as are 
\sii\ and \oiv. 

The decisive merit of Model $M1$ is to demonstrate that, 
with no extra free parameter, no change of shell size and 
no significant change of source SED, {\sl the `$r$(\oiii) problem' met 
in $N0$ can be essentially solved} just by considering radiation-bounded 
filaments embedded in a lower density (higher ionization) medium instead of 
a clumped shell at constant density. While the normalized $r$(\oiii) is 0.96 
(\oiii\ in Table~\ref{tab_compa}), no dense or distant clumps of the 
kind postulated by SS99 (Sect.~\ref{prev_IZ}) are needed to account for 
low-ionization lines. The pressure contrast is $<$~5. 

$Q_{\rm abs}$/$Q$ is again too small, but $Q$ cannot decrease because 
$\delta_4$ is close to unity (Table~\ref{tab_mod}). In $N0$, $\delta_4$ 
was (perhaps anomalously) small, enabling a shift from $N0$ to $N1$. 
Hardening the already hard primary ($\alpha$ = 0.7 
due to large $T1$; Fig.~\ref{fig_SED}) would enhance \ariv, 
predicted too strong. Also, $f^{cov}$ is only 0.20, resulting in an 
artificial cigar-like radial distribution, in which the low-density gas 
exactly shields the denser filaments from direct primary radiation. 

Obviously, the limits of the one-sector model are being reached. 
A matter-bounded sector is to be added for the sake of a larger 
absorbed fraction of photons in the shell, 
but not principally to improve the already quite 
satisfactory intensities of \oiii\la4363 and \oi\la6300. 

\subsection{Two-sector photoionization models: $M2$, $M3$, $M4$}
\label{res_M2comp}

The enhancement of \heii\ and \oiii\ related to the matter-bounded sector 
must be balanced by a weaker/softer SED. Models illustrate 
the influence of the SED (Fig.~\ref{fig_SED}). 

Given a SED and both covering factors, 
then $\tau_{\rm m}$(2), $\tau_{\rm c}$ and $\delta_4$ can be fine-tuned to 
account for I(\hb), \oiii/\oii\ and \heii. Iterations along the same lines as 
for Model $M1$ eventually lead to a model, provided that the limits on 
parameters are respected (Table~\ref{tab_conv}). A two-parameter 
model sequence can be attached to any SED by considering several pairs 
($f^{cov}_1$, $f^{cov}_2$), but little freedom is attached to $f^{cov}_1$, as 
Sector\,1 is where $\sim$\,75\% of \hb\ and most of \oii\ come from. 
Then $f^{cov}_1$ must be on the order of, or moderately larger 
than the $f^{cov}$ of the one-sector model, say, in the range 0.2$-$0.3. 
Also $f^{cov}_2$ cannot be small since one-sector models are rejected 
(Sect.~\ref{res_M1comp}) and $f^{cov}_1$\,+\,$f^{cov}_2$ must be kept 
significantly less than unity to enable \heii\ excitation beyond the 
shell (Sect.~\ref{mod_shell}), implying  $f^{cov}_2$ $\sim$ 0.3$-$0.6. 

\subsubsection{Model $M2$ and variants}
\label{res_M2}

Model $M2$ (Col.\,5 of Table~\ref{tab_mod}) is the first `complete' 
model. $Q_{\rm abs}$/$Q$ is at the low end of the nominal interval. 
Ouput of line intensities (Cols.\,5--6 of Table~\ref{tab_res}; 
Col.\,6 of Table~\ref{tab_compa}) is to be contrasted to the $N0$ 
output. Runs of \Ne\ and \Te\ 
with $R$ are shown in  Figs.~\ref{fig_Ne}c--\ref{fig_Te}c. The sharp 
`spike' of the \Ne\ curve shows how thin a radiation-bounded filament is 
compared to the shell. Runs of \Ne\ and \Te\ are best 
seen in plots {\sl versus} $\tau$ (Figs.~\ref{fig_Ne}d--\ref{fig_Te}d). In 
plots for $M2$, vertical arrows mark the outer boundary of Sector~2. 
Ion fractions O$^{n+}$/O {\sl versus} $R$ and $\tau$ are shown 
in Figs.~\ref{fig_Oxy}c$-$\ref{fig_Oxy}d. 

Sufficient ionization is maintained in $M2$ owing to the lower average 
density, which also helps increasing \Te\ in the high-ionization layers, 
despite the significantly softer radiation field (smaller $Q_{\rm He}$/$Q$ 
and larger $\alpha$, Fig.~\ref{fig_SED}b): the inner \Te\ is now 
2.5$\times$10$^4$\,K. In Fig.~\ref{fig_Te}d, the jumps of \Te\ at 
$\tau$\,$\sim$\,0.1 and $\sim$\,4.7 correspond to the boundaries 
of the He$^{2+}$ shell (fairly well traced by O$^{3+}$  
in Fig.~\ref{fig_Oxy}d) and the filament respectively. 
Comparing Fig.~\ref{fig_Oxy}d to Fig.~\ref{fig_Oxy}b, 
the ionic fractions are qualitatively similar in Model $M2$ and 
Run $N1$, but the transition from O$^{2+}$ 
to O$^{+}$ is sharper and occurs at a smaller optical depth in $M2$. 

Average properties and abundances of Model $M2$ are quite similar to those 
of $M1$ and the predicted \oiii\la4363 intensity is again slighly weak, 
although the score of $M2$ is significantly better for \ariv\ and \siv\ 
(Col.\,6 of Table~\ref{tab_compa}). 

Variants to Model $M2$ can be obtained by changing $f^{cov}_1$ and 
$f^{cov}_2$ within limits, while retaining source 
parameters (except for minute fine-tuning of $\delta_4$). 
In Col.\,1 of Table~\ref{tab_vari} are listed 7 shell parameters 
and 6 lines extracted from Table~\ref{tab_compa}. 
$M2$ (Col.\,2 of Table~\ref{tab_vari}) is compared to 
re-converged models $M2b$ (Col.\,3) and $M2c$ (Col.\,4). 
Increasing $f^{cov}_2$ from a small to a large value, with $f^{cov}_1$ 
left unchanged, structure parameters are not much changed 
except for a decrease of $\tau_{\rm m}$(2) 
and a small decrease of O/H due to the larger weight of the hot 
high-ionization zone. Accordingly, \ariv\ is increased, 
but \oiii\la4363 is increased by only 1\%. Decreasing $f^{cov}_1$ from 
0.26 to 0.22, \hb\ is recovered by increasing $\tau_{\rm m}$(2) 
and \oii\la3727 by decreasing $\tau_{\rm c}$, with the consequence 
that $P_{\rm in}$ must decrease, thus \Te\ increases and O/H 
decreases. Finally the \la4363 intensity increases up to 
the observed value, \ariv\la4740 and \siv\la10.5$\mu$ 
increase and \oi\la6300 decreases. 

\begin{table}
%Table 5
\caption[]{Influence of shell parameters on selected line intensity predictions for models $M2$ and $M4$}
\begin{tabular}{l|cccccc}
\hline\hline
 Param. / line         &  M2  &  M2b &  M2c &  M4b &  M4  &  M4c \\
                       &   2  &   3  &   4  &   5  &   6  &   7  \\
\cline{1-7}
$f^{cov}_1$            & 0.26 & 0.26 & 0.22 & 0.29 & 0.29 & 0.22 \\
$f^{cov}_2$            & 0.30 & 0.60 & 0.60 & 0.30 & 0.60 & 0.60 \\ 
$\tau_{\rm m}$(2)      & 1.21 & 0.62 & 0.92 & 2.17 & 1.01 & 1.57 \\ 
$\tau_{\rm c}$         & 4.9  & 4.7  &  3.8 & 4.6  & 4.3  & 3.0  \\
$P_{\rm in}/k$/10$^5$  & 3.4  & 3.4  &  3.1 & 2.7  & 2.7  & 2.4  \\
$P_{\rm out}/k$/10$^5$ & 21.7 & 22.5 & 24.6 & 26.8 & 26.8 & 26.8 \\
O/H$\times$10$^7$      & 168. & 165. & 162. & 178. & 173. & 171. \\ 
\cline{1-7}
 {\oi}$\:$6300+        & 1.01 & 1.00 & 0.92 & 1.42 & 1.37 & 1.06 \\
 {\oiii}$\:$4363       & 0.96 & 0.97 & 1.00 & 0.92 & 0.94 & 0.95 \\
 {\oiv}$\:$25.9$\mu$   & 0.97 & 0.91 & 0.92 & 1.05 & 0.98 & 0.99 \\
 {\sii}$\:$6716        & 0.78 & 0.81 & 0.79 & 1.05 & 1.10 & 0.84 \\
 {\siv}$\:$10.5$\mu$   & 1.93 & 1.97 & 2.05 & 2.07 & 2.15 & 1.92 \\
 {\ariv}$\:$4740       & 1.39 & 1.48 & 1.62 & 1.55 & 1.77 & 1.81 \\
[0.1cm]
\hline
\end{tabular}
 
\label{tab_vari}
\end{table}

\subsubsection{Models $M3$ and $M4$}
\label{res_M34}

In Model $M3$, $T2$/10$^4$\,K is enhanced from 4 to 5. 
$Q_{\rm abs}$/$Q$ is larger than in $M2$ due to lower 
luminosity. The larger $T2$ increases the average energy of 
photons absorbed in the O$^{2+}$ region and the intensity of \oiii\la4363 
is slightly larger. In the selected example, \oiii\ is again exactly 
matched as in $M2c$, but for more `standard' $f^{cov}_1$ and $f^{cov}_2$ 
(Col.\,6 of Table~\ref{tab_mod}). Line intensity 
predictions are slightly improved (Col.\,7 of Table~\ref{tab_compa} 
{\sl versus} Col.\,4 of Table~\ref{tab_vari}). 

In Model $M4$, $T2$ is again as in $M2$, while $T1$/10$^4$\,K 
is enhanced from 8 to 12 and $L1$/$L2$ is halved. $Q_{\rm abs}$/$Q$ is 
close to its allowed maximum due to radiation hardening, which also leads 
to $\delta_4$\,$\ll$\,1 (Col.\,7 of Table~\ref{tab_mod}). The large flux 
just below 4\,ryd enhances simultaneously the high and low ionization lines, 
but the heating of the O$^{2+}$ region is lesser and the large $T1$ 
(positive curvature of the SED) does not favour a large $r$(\oiii) 
(Col.\,8 of Table~\ref{tab_compa}). From Table~\ref{tab_vari}, variants 
$M4b$ (Col.\,5) and $M4c$ (Col.\,7) of $M4$ (Col.\,6) fail to enhance 
\oiii\la4363 up to the observed value. Increasing the source luminosity 
by 20\%, thus decreasing $Q_{\rm abs}$/$Q$ from 0.43 to 0.36, has strictly 
no effect on the predicted \oiii\ after convergence.

\section{Discussion}
\label{disc}

Irrespective of the `technical' demand raised by $Q_{\rm abs}$/$Q$ in 
Sect.~\ref{res_M1comp}, a two-sector model is the minimum complexity 
of any shell topology (Sect.~\ref{mod_shell}). The two degrees 
of freedom attached to the matter-bounded sector are inescapable. 

Model $M4$ (and variants) appears slightly less successful than $M2$ and 
$M3$ concerning \oiii\la4363 and the high-ionization lines (\ariv\la4740, 
\siv\la10.5$\mu$). $M4$ has a possibly less likely SED and presents the 
largest $P_{\rm out}$/$P_{\rm in}$. The discussion focuses on Models $M2$ 
and $M3$, with $M2$ the `standard' from which variants are built. 

\subsection{Spectral energy distribution}
\label{disc_SED}

Accounting for a $Q_{\rm abs}$/$Q$ larger than, say, 1/3 turns out 
to be demanding. Selected models correspond to nearly maximum 
possible values for each SED. Acceptable $Q_{\rm abs}$/$Q$ can indeed be 
obtained, but the latitude on the SED and power of the ionizing source 
is narrow. The uncertainties in evolutionary synthetic cluster models 
(Appendix\,\ref{WRstars}) and in the evolutionnary status of \IZ\,NW itself 
are sufficient to provisionally accept the `empirical' SED corresponding 
to preferred models $M2$ or $M3$ (Fig.~\ref{fig_SED}) as plausible. 
The `predicted' typical trend is 
$L_{\nu}$($h\nu$\,=\,1\,$\rightarrow$\,4\,ryd) $\propto$ exp($-h\nu$/ryd). 

\subsection{Ionized gas distribution}
\label{disc_gas}

The model is most specific in that emission lines partly arise from a low 
density gas, while the largest \Ne\ is \Ne(\sii). A density $\leq$~10\,\cc\ 
(Table~\ref{tab_mod}) appears very low by current standards of photoionization 
models for BCDs (Sect.~\ref{prev_indiv}). Nevertheless, the superbubble model 
of Martin (1996) is consistent with a current SFR = 0.02\,\Msun\,yr$^{-1}$ for 
the whole NW+SE complex. The two best estimates in the compilation by Wu07 are 
0.03 and 0.02\,\Msun\,yr$^{-1}$. Adopting half the Martin (1996)'s rate and a 
wind injection radius of 0.1\,kpc (1.5\arcsec\ at 13\,Mpc) for the NW cluster 
alone, expression (10) in Veilleux et al. (2005) suggests an inner pressure 
of the coronal gas $P/k$ $\sim$ 3$\times$10$^{5}$\,K\cc, hence an ambient ISM 
number density $\sim$~12\cc\ in the inner region 
(\Te $\sim$ 2.5$\times$10$^{4}$\,K; Sect.~\ref{res}), or else 
\Nh\ $\sim$ 12/2.3 = 5\cc\ for a photoionized gas in pressure balance with 
the coronal phase which presumably permeats the shell. Although $\epsilon$ is 
unity in models, this phase can fill in the volume corresponding to Sector\,3. 

\subsection{Optical and UV lines}
\label{disc_opt}

\oiii\ lines are discussed in Sects.~\ref{disc_ab} \& \ref{disc_eva}. 
\civ\ and \sit\ are accounted for within uncertainties. 
Other UV lines are elusive (Col.\,6 of Table~\ref{tab_res}). 

Computed fluxes for 8 optical lines are within 20\% of observation 
(Table~\ref{tab_compa}), which is satisfactory considering the weakness 
of some of the lines. The 10$-$15\% discrepancy on the ratio 
\siii\,6312/9531 does not challenge the model itself, given the 
various uncertainties. The $\sim$\,20\% underestimation of \sii\ should 
be considered with respect to \siii. The line \la9531 is matched, 
but the far-red flux may be less reliable, and \la6312 departs 
from observation about in the same way as \sii, but \la6312 
is a weak line. The exact status of \sii\ is undecided. 

The \Ne-sensitive intensity ratio \feiii\la4986/\la4658 is 
somewhat small in $N0$, large in $N1$ and more nearly 
correct in $Mi$ models. Would \sii\ be emitted in a 
high-\Ne\ gas component as suggested by SS99 and V02, then \feiii\,4986, 
roughly co-extensive with \oii, would be undetectable. The weak line 
\feiv\,4906, co-extensive with \oiii, confirms the ion distribution 
of the $Mi$ models and the iron abundance, although the 
agreement with observation is partly fortuitous. \fev\,4227 is 
overpredicted by a factor $\sim$3, but the observed intensity 
is very uncertain and could be 2--3 times stronger than the 
quoted value, as judged from published tracing (TI05). Also, only 
one computation of collision strengths has ever been done for the 
optical lines of the difficult \fev\ ion (Appendix\,\ref{fiat_misc}). 
Finally, the ionic fraction Fe$^{4+}$/Fe, less than 5\%, is subject 
to ionization balance inaccuracy. The predicted intensity 
$\sim$~I(\la4227\AA)/4 of \fev\,4071 enhances the computed flux of 
\sii\,4068 up to the observed value (Tables~\ref{tab_res} \& \ref{tab_compa}). 

The observed \hei\ line intensities are inconsistent (Table~\ref{tab_res}), 
due to stellar lines (Sect.~\ref{mod_constraints}). \ariv\,4711, blended 
with \hei\,4713, is therefore useless. The weak \ariv\,4740 tends to be 
overestimated by $\sim$~50\% in the preferred models. Trial calculations 
show that, adopting a recombination coefficient 12 times the radiative one 
(instead of 8 times, Appendix\,\ref{fiat_misc}) and dividing Ar/H by 1.13, 
\ariii\ and \ariv\ would be matched in $M2$. 

\subsection{Infrared fine-structure lines}
\label{disc_ir}

\subsubsection{\neiii\ and \siii}
\label{disc_ir_nesiii}

The reliably observed IR lines with optical 
counterparts, \neiii\,15.5$\mu$ and \siii\,18.7$\mu$, are 
very well matched, confirming the scaling adopted for the 
{\sc spitzer} fluxes and the model temperatures. 
The $Mi$ models, globally hotter, are more successful than 
the $Ni$ runs. No $t^2$ in excess of the one of the adopted 
configuration (Fig.~\ref{fig_t2}d) is required. 

The predicted intensity of \siii\,33.5$\mu$ is only 40\% of the observed 
value. Since the theoretical ratio of the \siii\ IR lines is insensitive 
to conditions in \IZ, looking for alternative models is hopeless. 
The collision strengths $\Omega$ for the \siii\ lines may not be of ultimate 
accuracy, as the results of Tayal \& Gupta (1999) and 
Galavis et al. (1995) differ, but the more recent $\Omega$'s are likely 
more accurate. Also, the predicted \siii\,33.5$\mu$ is even worse 
using older data. 
Since Wu07 cast doubts on the accuracy of the flux calibration at the 
end of the {\sc spitzer} spectrum, it is assumed that the 
\siii\ atomic data are accurate and that the observed fluxes around 
\la34$\mu$ should be divided by 2.3. 

\subsubsection{\oiv}
\label{disc_ir_oiv}

If the drift of flux calibration at \la34$\mu$ 
(Sect.~\ref{disc_ir_nesiii}) smoothly vanishes towards shorter wavelengths, 
the \oiv\,26$\mu$ flux may still be overestimated. Conversely, the 
{\sc spitzer} field of view encompasses \IZ\,SE, which emits little \heii, 
leading to underestimate \oiv/\hb\ in \IZ\,NW. Since these effects act 
in opposite directions, the original \oiv/\hb\ is adopted for \IZ\,NW. 

As shown in Fig.~\ref{fig_Oxy}, O$^{3+}$ and O$^{2+}$ coexist in the He$^{2+}$ 
zone. O$^{3+}$/O$^{2+}$ and therefore \oiv\la25.88$\mu$/\heii\la4686 
as well are sensitive to \Ne. In $N0$, \heii\ is matched and \oiv\ is 
strongly underpredicted (Table~\ref{tab_compa}). The predicted \oiv\ flux 
improves in the conditions of $N1$ and even by-pass 
observation in $M1$, whose ionizing flux is however 
too large (Sect.~\ref{res_M1comp}). In the standard Model $M2$, the predicted 
\oiv\ exactly matches observation after adding the blended line 
\fev\,25.91$\mu$, whose computed flux is $\sim$~2\% of \oiv. Since relevant 
atomic data are reliable, {\sl \oiv\,25.88$\mu$ indicates that \Ne\ must be 
on the order of 10\cc\ in the \heii\ emitting region of the \IZ\,NW shell}. 
The model density results from general assumptions (photoionization by stars, 
shell geometry, $\epsilon$~=~1, Eq.~1, etc.) and a requirement to match a few 
basic line intensities with no reference to  high-ionization lines, but \heii. 
The computed \oiv\ intensity is a true prediction, especially as the models 
were essentially worked out prior to IR observations: the spectrum presented 
by Wu et al. (2006) showed the predicted \oiv\ line, finally noted by Wu07. 

\subsubsection{\siv}
\label{disc_ir_siv}

The predicted \siv\,10.5$\mu$ flux is twice the observed one. 
The collision strengths obtained by Tayal (2000) and 
Saraph \& Storey (1999) for this line are in good agreement. The 
average fractional concentration of S$^{3+}$, $\sim$~1/3, is stable 
in different models because sulfur is mostly 
distributed among the three ions S$^{2+}$$-$S$^{4+}$. Displacing the 
ionization balance by changing, \eg, the gas density tends to make either 
S$^{2+}$ or S$^{4+}$ to migrate to S$^{3+}$. Only in the unsatisfactory 
run $N0$ is \siv\ accounted for. 

Two-sector, constant-pressure `models' allowing $\epsilon$\,$<$\,1 and 
using the SED of Model $M2$ were run with the conditions 
$Q_{\rm abs}$/$Q$ $>$ 0.3 and O/H $<$ 1.7$\times$10$^{-5}$. 
In these trials, the \oiii\,5007 and $r$(\sii) constraints 
(Table~\ref{tab_conv}) are relaxed and the observed \siv/\siii\ ratio 
is exactly matched by playing with \Ne\ (gas pressure) and $\epsilon$. 
Despite ample freedom and because of the higher \Ne~$\sim$~25\cc, the computed 
\oiv\ flux is at most 60\% of the observed one. Thus, forgetting 
other difficulties, the suggestion 
is that the excess \siv\ flux can only be cured at the expense of \oiv. 
A broader exploration of the SED (discontinuities?) and the gas distribution 
could be undertaken. 

The \siv\,10.5$\mu$ flux published by Wu et al. (2006) was 25\% larger than 
according to Wu07. The new value should be preferred, but this difference 
is at least indicative of possible uncertainties. The ratio \siv/\hb\ may 
also be intrinsically larger in \IZ\,NW than in \IZ\,SE. 

The theoretical ionization balance of some ions of sulfur (and argon) is 
subject to uncertainties (Appendix\,\ref{fiat_misc}). The observed \siv/\siii\ 
ratio can be recovered in $M2$ if the S$^{2+}$ recombination coefficient is 
multiplied by a factor 2.3, which is perhaps too large a correction: 
then, both \siii\ and \siv\ are matched if S/H is divided by 1.33, 
with the caveat that the predicted \sii\ intensities are divided by 1.3. 
A combination of observational and theoretical effects just listed 
could alleviate the `\siv\ problem'. 

\subsection{Low ionization fine-structure lines}
\label{disc_ir_low}

Although the error bars of order 10\% quoted by Wu07 may not 
include all sources of uncertainties, both \neii\,12.78$\mu$ 
and \feii\,25.98$\mu$ are detected in high-resolution mode. 
\sid\,34.80$\mu$ is strong, even though the flux quoted by Wu07 
(Table~\ref{tab_compa}) may be too large (Sect.~\ref{disc_ir_nesiii}). 
Usually, the bulks of \sid\,35$\mu$ and \feii\,26$\mu$ arise from a 
Photon-Dominated Region (PDR), at the warm \hi\ interface between an 
ionization front and a molecular cloud (\eg, Kaufman et al. 2006). 
Schematically in a PDR, the photo-electric heating by UV radiation 
on dust grains (and other molecular processes) is balanced by 
fine-structure (and molecular) line emission. The small reddening intrinsic 
to \IZ\ (Sect.~\ref{abs_red}) and the `large' gaseous iron content 
(Sect.~\ref{disc_ab}) imply that little dust is available. 
Molecules and PAHs are not detected in \IZ\ 
(Vidal-Madjar et al. 2000; Leroy et al. 2007; Wu07). 
The classical PDR concept may therefore not apply to \IZ, raising 
the question of the origin of \sid\,35$\mu$ and \feii\,26$\mu$, both 
underpredicted by factors 5--10 in the models (Table~\ref{tab_compa}). 

\begin{figure}
%Fig.6
\resizebox{\hsize}{!}{\includegraphics{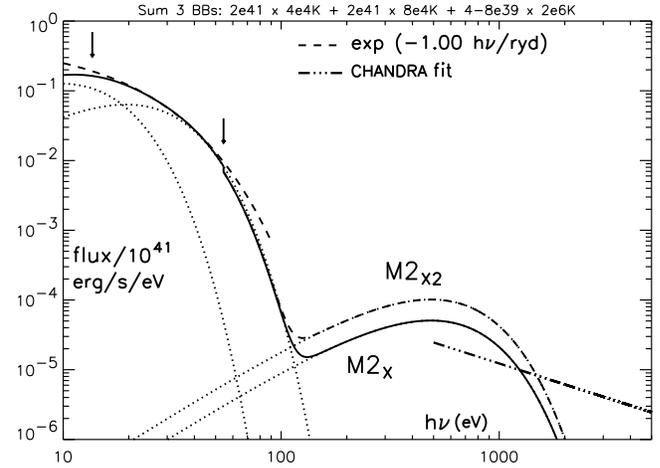}}
\caption[]{
As in Fig.~\ref{fig_SED}b for the variants of $M2$ comprising a hot black body emitting at soft X-ray energies: Model $M2_{\rm X}$ (solid line) and $M2_{\rm X2}$ (dash-dotted line). The unit flux on the ordinate is now 10$^{41}$\ergs\,eV$^{-1}$. The dash-triple-dotted line is one possible fit to the {\sc chandra} data for \IZ\,NW. 
}
\label{fig_SEDX}
\end{figure}

A way to produce a `pseudo-PDR' 
is X-ray heating. Two new variants of $M2$ are considered, in which a 
hot black body representing a soft X-ray emission from \IZ\,NW is added 
to the original SED. The adopted temperature is \TX\ = 2$\times$10$^{6}$\,K 
and the luminosity \LX\ = 4 and 8$\times$10$^{39}$\ergs\ for variants 
$M2_{\rm X}$ and $M2_{\rm X2}$ respectively (Fig.~\ref{fig_SEDX}). 
The actual X-ray luminosity of \IZ, $\sim$ 1.6$\times$10$^{39}$\ergs\ 
in the 0.5--10 kev range of {\sc chandra}, mainly arises from the centre 
of \IZ\,NW and is consistent with a power law 
of slope $-1$ (Thuan et al. 2004), drawn in Fig.~\ref{fig_SEDX}. The SED 
for $M2_{\rm X}$ is a relatively high, yet plausible extrapolation of the 
{\sc chandra} data. $M2_{\rm X2}$ is considered for comparison purpose.  
\Nh, still governed by Eq.~(1), is leveled out at 300\cc. 

\begin{figure}
%Fig.7
\resizebox{\hsize}{!}{\includegraphics{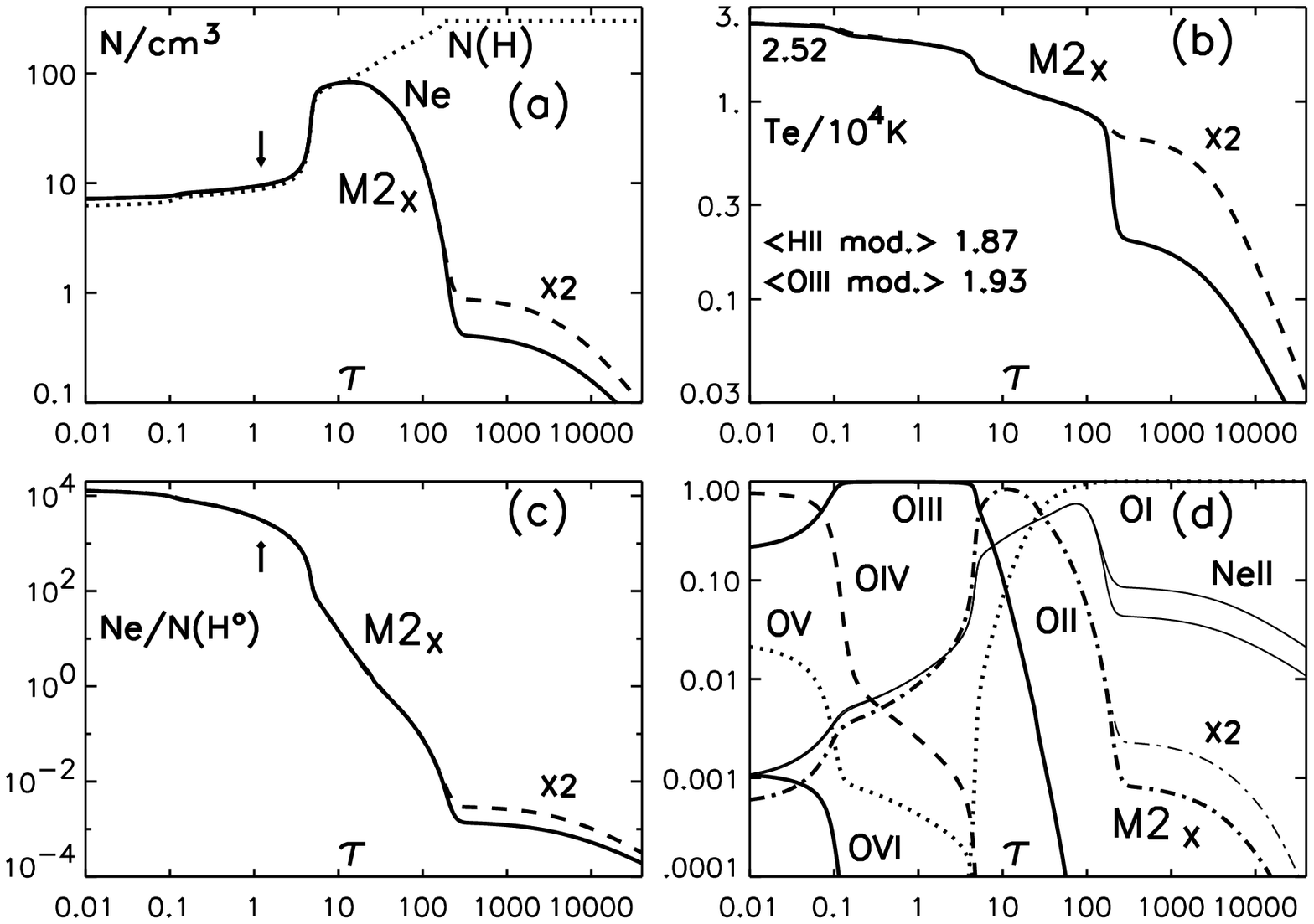}}
\caption[]{
Model $M2_{\rm X}$ for \IZ\,NW. ($a$) Electron density \Ne\ (solid line) and total hydrogen density N(H) (dotted line) {\sl versus} optical depth $\tau$ at 1~ryd: N(H) levels out at 300\cc. ($b$) Electron temperature \Te/10$^4$\,K: the maximum value and average values weighted by \Ne$\times$N(H$^+$) and by \Ne$\times$N(O$^{2+}$) are noted for the full 2-sector model ($<$~mod.~$>$). ($c$) Ratio of \Ne\ to atomic hydrogen density N(H$^0$). In panels ($a$)--($c$), the dashed lines correspond to $M2_{\rm X2}$. ($d$) Local fractional concentrations of O$^{0}$ (dotted line), O$^{+}$ (dot-dashed line), O$^{2+}$ (solid line), O$^{3+}$ (dashed line), O$^{4+}$ (dotted line again) and O$^{5+}$ (solid line again): the thinner dash-dotted line ($\tau$ $>$ 200) corresponds to O$^+$ in $M2_{\rm X2}$ and the thinner solid line (split into $M2_{\rm X}$ and $M2_{\rm X2}$ for $\tau$ $>$ 200) to Ne$^+$. 
}        
\label{fig_M2X}
\end{figure}

The run of physical conditions with $\tau$ is shown in 
Fig.~\ref{fig_M2X} for $M2_{\rm X}$. While $<$\Te$>$ is increased by only 
$\sim$~100\,K in the \hii\ region, a warm low-ionization layer develops 
beyond the ionization front. Computation is stopped at \Te\ = 100\,K 
($\tau$ $\sim$ 7$\times$10$^{4}$, compared to 250 in $M2$). In $M2_{\rm X2}$ 
(dashed lines in Figs.~\ref{fig_M2X}a--\ref{fig_M2X}c), the new layer is 
hotter and more ionized (final $\tau$ $\sim$ 1.2$\times$10$^{5}$). 
The geometrical thickness of the \hi\ layer is 21\% and 35\% of the 
\hii\ shell in $M2_{\rm X}$ and $M2_{\rm X2}$ respectively. In the 
\hi\ zone, O$^+$/O does not exceed $\sim$~10$^{-3}$, in marked contrast 
with Ne$^+$/Ne, overplotted as a thin solid line in Fig.~\ref{fig_M2X}d. 
Lines \neii\,12.8$\mu$ and \arii\,7.0$\mu$ are usually discarded in PDR 
models on the basis that the ionization limits of Ne$^0$ and Ar$^0$ exceed 
1\,ryd. Here, owing to the scarcity of free electrons and the lack of 
charge exchange with H$^0$, photoionization by soft X-rays can keep 
1--10\% of these elements ionized. 

In \hi\ regions, cooling is due to inelastic collisions with H$^0$. Reliable 
collisional rates exist for the main coolents \cii\,157$\mu$ 
(Barinovs et al. 2005) 
and \oi\,63$\mu$ (Abrahamsson et al. 2007; several processes need be 
considered for \oi: see Chambaud et al. 1980; P\'equignot 1990) and 
for \sid\,35$\mu$ (Barinovs et al. 2005), but not for, \eg, \feii\,26$\mu$. 
Following Kaufman et al. (2006), it is assumed that the cross-section for 
H$^0$ + Si$^+$ also applies to fine-structure transitions of other singly 
ionized species. Concerning \feii\ (ground state $^6$D$_{9/2}$), 
collisions to $^6$D$_{7/2}$ follow the above rule, but cross-sections for 
transitions to the next $^6$D$_{J}$ are taken as 2/3, 2/4, etc. of the 
first one. 

\begin{table}
%\begin{table*}
%Table 6
\caption[]{Fine-structure line intensities$^a$ and X-rays}
\begin{tabular}{lccccccc}
\hline\hline
 Line$^b$  & Obs. & \multicolumn{2}{c}{$M2$} & \multicolumn{2}{c}{$M2_{\rm X}$} & \multicolumn{2}{c}{$M2_{\rm X2}$} \\
\cline{1-8}
\hi\ colls.$^c$ &   -  &  no  &  yes &   no & yes &   no & yes \\
{\cii}$\:$158.  &   -  &  2.1 &  2.1 &  124 & 119 &  224 & 204 \\
{\oi}$\:$63.2   &   -  &  4.7 &  4.6 &  274 & 210 &  624 & 450 \\
{\neii}12.8     &   9  &  1.9 &  3.6 &  1.9 & 5.5 &  2.0 &  16 \\
{\sid}$\:$34.8  &157$^d$&  21 &   22 &   37 &  80 &   55 & 190 \\
{\arii}$\:$7.0  &   -  &  1.2 &  1.3 &  1.4 & 2.2 &  2.2 & 7.9 \\
{\feii}$\:$26.0 &  34  &  3.4 &  4.8 &  5.8 &  38 &  9.3 & 108 \\
{\feii}$\:$35.3 &   -  &  0.8 &  1.4 &  1.1 & 9.5 &  1.7 &  33 \\
[0.1cm]
\hline
\end{tabular}

\ \ $^a$ In units \hb\ = 1000. \\
\ \ $^b$ Wavelengths in $\mu$m. \\
\ \ $^c$ Inelastic collisions of \sid, \feii, \neii\ and \arii\ with \hi\ omitted (`no') or included (`yes'). \\
\ \ $^d$ The re-calibrated \sid\ flux is 68 after Sect.~\ref{disc_ir_oiv}. \\
 
\label{tab_fs}
\end{table}

Low-ionization IR lines, including dominant coolents and 
other unobserved lines, are considered in Table~\ref{tab_fs}. 
Line identifications and observed intensities appear in Cols.\,1 \& 2. 
The results of two computations are provided for $M2$ (Cols.\,3-4), 
$M2_{\rm X}$ (Cols.\,5-6) and $M2_{\rm X2}$ (Cols.\,7-8). In the first one, 
the excitations of \sid, \feii, \neii, and \arii\ (but not \oi\ and \cii) 
by collisions with \hi\ are inhibited. In the second one, they are included 
according to the above prescription. Comparing different odd columns (`no') 
of Table~\ref{tab_fs}, the rise of line intensities as the 
\hi\ zone develops shows that the excitation of, \eg, \sid\ by free 
electrons is still active. \neii\ is stable, due to the 
scarcity of Ne$^+$ (Fig.~\ref{fig_M2X}d) relative to Si$^+$. 
Comparing now odd columns to even columns, 
it is seen that \hi\ collisions, ineffective in the `normal' \hii\ 
region model $M2$ (except, quite interestingly, for \neii), strongly 
enhance the excitation rates. After subtracting emission from the 
\hii\ region ($M2$), \hi\ collisions contribute 75--80\% of 
the excitation (virtually 100\% in the case of \neii). 

Concerning \sid\,35$\mu$, the atomic data are not controversial and the 
re-calibrated flux of 68 is reasonably well defined (Sect.~\ref{disc_ir_oiv}). 
Then $M2_{\rm X2}$ (Col.\,8 of Table~\ref{tab_fs}) is excluded, while 
$M2_{\rm X}$ (Col.\,6) or an even weaker soft X-ray source 
(closer to the {\sc chandra} extrapolation) can account 
for \sid. Although the remarkably coherent predictions for \sid\,35$\mu$ 
and \feii\,26$\mu$ are partly fortuitous, they are consistent with 
(1) the \feii\ collision strength is correctly guessed, 
(2) Fe/Si is the same in the ionized and neutral gas, and 
(3) the excitation by soft X-ray heating is viable. 
Concerning \neii\,12.8$\mu$, the discrepancy with observation (factor 0.6) 
may not be significant, as the line is weak and its detection 
in the {\sl low}-resolution mode is not taken as certain by Wu07. 
The collision strength may be too small. 

Summarizing, a plausible extrapolation to soft X-rays of 
the {\sc chandra} flux can provide an explanation 
to the relatively large intensity of \sid\,35$\mu$ and other 
fine-structure lines in \IZ. This is considerable support 
to the general picture of photoionization as the overwhelmingly dominant 
cause of heating of the \hii\ region, since {\sl heating by conversion of 
mechanical energy appears unnecessary even in regions protected from 
ionization and heating by star radiation}. Full confirmation should await 
reliable collisional excitation rates by \hi\ for fine-structure lines 
of {\sl all} singly ionized species. The soft X-rays from \IZ\,NW have 
little effect on the \hii\ region: both \oi\,6300 and \oiv\,25.9$\mu$ 
are increased by 3\% and \fevi\ by 30\%. The 1.7\% enhancement 
of \oiii\,4363 is of interest (Sect.~\ref{disc_eva}). 

\subsection{Stability of results}
\label{disc_stab}

The relative stability of the predicted line intensities is a consequence 
of the set of constraints (Table~\ref{tab_conv}). Allowing for a range of 
values, a broader variety of results could be obtained. Are conclusions 
dependent on input data? 

The only basic line showing substantial variability in 
different spectroscopic studies is \oii\,3727. This line is 
sometimes found to be stronger than the adopted value 
(VI98; TI05). 
In a new variant $M2v$ of Model $M2$, 
the observed \oii\ intensity is assumed to be 20\% larger than in 
Tables~\ref{tab_res} \& \ref{tab_compa} and the covering factors are 
left unchanged. The inevitable 18\% increase of the already too strong 
line \oii\,7320+30 is not too significant, considering that the 
\la7325 flux is very uncertain and may not correspond to a 
slit position with stronger \la3727. Owing to the larger fractional 
abundance of O$^+$, O/H is increased by 5\%. The larger weight of 
low-ionization layers induces a 3\% increase of Ne/H 
and a 13\% decrease of N/H and Fe/H, since the \nii\ and \feiii\ intensities 
were left unchanged. Both \ariv\ and \siv\ decrease by a few \%, whilst \oiv\ 
increases by 6\% and both \sii\ and \oi\ increase by $\sim$~11\%. Ar/H and 
the \siii\ lines increase by only 1\% and \oiii\la4363 {\sl is unchanged}. 

In a more extreme example, $M2cv_{\rm X}$, with an assumed \oii\ intensity 
of 322 instead of 238 (factor 1.35), the $f$'s as in $M2c$ and 
the SED as in $M2_{\rm X}$ (reconverged $\tau_{\rm c}$ = 2.5 and 
$P_{\rm in}$/$P_{\rm out}$ = 9), \oiii\ is exactly matched again, 
\oi\ is +2\% off, \oiv\ +6\%, \siv\ +87\% and \sii\ only $-$8\%. 

Thus, changes are moderate and turn out to alleviate difficulties noted 
in Sects.~\ref{disc_opt} \& \ref{disc_ir}, \eg, the weakness of the 
\sii\ doublet. The computed $r$\oiii\ is robust. 

\subsection{Elemental abundances}
\label{disc_ab}

To first order, O/H reflects $<$\Te(O$^{2+}$)$>$, 
related to $r$(\oiii), \ie, the predicted \oiii\la4363 intensity. 
Models $M3$ and $M2c$ both almost exactly fit \la4363 and share the 
same O/H = 1.62$\times$10$^{-5}$, which is the best estimate, provided that 
(1) oxygen lines have been given optimal observed intensities, 
(2) these models faithfully represent the \hii\ region, 
and (3) collision strengths are accurate. 

Concerning line intensities, \oiii\,5007 is quite stable 
in different spectra of \IZ\,NW and the reasonably large, 
yet representative, ratio $r$(\oiii) (Sect.~\ref{spectr}) 
is taken for granted, since our objective is deciding 
whether this specific ratio can be consistently explained assuming 
photoionization by stars. In model $M2cv_{\rm X}$ (Sect.~\ref{disc_stab}), 
which also fits exactly the \oiii\ lines, \oii\,3727 was assumed to be 
enhanced by 35\%, leading to O/H = 1.74$\times$10$^{-5}$. 

\begin{figure}
%Fig.8
\resizebox{\hsize}{!}{\includegraphics{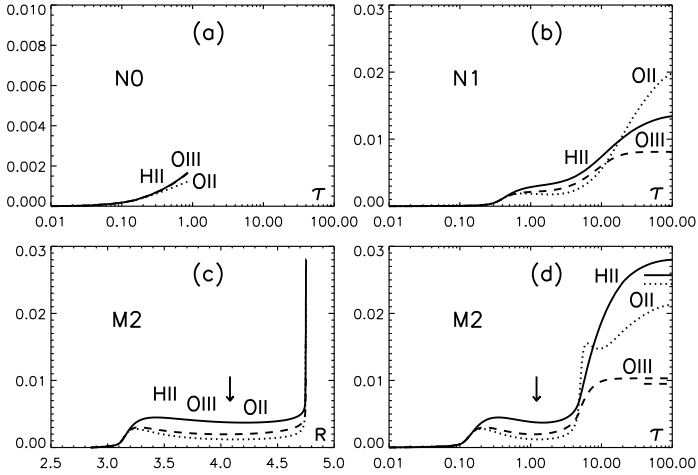}}
\caption[]{
Cumulative mean squared relative temperature fluctuations $t^2$ (Peimbert, 1967) weighted by \Ne$\times$N(H$^+$) (solid line), \Ne$\times$N(O$^+$) (dotted line) and \Ne$\times$N(O$^{2+}$) (dashed line) {\sl versus} $\tau$ in \IZ\,NW models: (a) $N0$, (b) $N1$, (d) $M2$; and {\sl versus} $R$: (c) $M2$. Note the expanded vertical scale in panel (a). The short horizontal bars to the right of panel (d) correspond to the means over the full model. Vertical arrows as in Fig.~\ref{fig_Ne}. 
}
\label{fig_t2}
\end{figure}

Concerning models, 
the difference between the \Te\ directly derived from $r$(\oiii), 
\Te(\oiii) = 19\,850\,K, and the \Ne$\times$$N$(O$^{2+}$) weighted average, 
$<$\Te(O$^{2+}$)$>$ = 19\,650\,K, corresponds to a formal 
$t^2$(\oiii) = 0.012, similar to the computed $t^2$(O$^{2+}$) = 0.010. 
This difference makes only 1\% difference for O$^{2+}$/H$^{+}$ 
and an empirical estimate neglecting $t^2$ should nearly coincide 
with model results for this ion. A major feature of the \Te\ profile 
is that the difference $<$\Te(O$^{2+}$)$>$ -- $<$\Te(O$^{+}$)$>$, which was 
only 300\,K in $N0$ and 3\,200\,K in $N1$, is 6\,600\,K in best models. 
Models are essential in providing a \Te\ to derive O$^{+}$/H$^{+}$ from 
\oii\,3727, as \Te(\oii) is poorly determined from the uncertain 
\oii\,7325 and $t^2$(\oii) is inaccessible. 

A `canonical' O/H for \IZ\,NW is 1.46$\times$10$^{-5}$ (SK93, ICF99),
11\% less than the present 1.62$\times$10$^{-5}$, out of which 4\% are due to 
collisional excitation of \hb\ and the remaining 7\% could be a non-trivial 
consequence of the relatively large $t^2$'s obtained for some ions in the 
present models (Table~\ref{tab_mod}; Fig.~\ref{fig_t2}), although 
differences in collision strengths may also intervene at the 2\% level 
(Appendix\,\ref{fiat_oiii}). 

The silicium and sulfur abundances were not fine-tuned in models. The 
computed \sit\ flux suggests dividing the assumed Si/H by 1.25. 
Concerning S/H, \sii\ is underestimated, \siii\ globally 
underestimated and \siv\ neatly overestimated. Correcting the ionization 
balance \siv/\siii, S/H should be divided by 1.3 in best models 
(Sect.~\ref{disc_ir_siv}), but \sii\ is then underestimated. Since a 
combination of effects may explain the overestimation of \siv, the model S/H 
is tentatively divided by 1.15. Similarly, \ariv/\ariii\ is best accounted for 
if Ar/H is divided by 1.13 (Sect.~\ref{disc_opt}), but the \ariv\ line is weak 
and the adopted correction factor is 1.06. Thus, S/Ar in \IZ\ is within 
10\% of the solar value, in agreement with a 
conclusion of Stevenson et al. (1993). The iron abundance relies 
on the \feiii\ lines, since \feii\,26$\mu$ does not arise from the 
\hii\ region, while the \feiv\ and \fev\ intensities are uncertain. 
The \feiii\ intensities are from a spectrum in which \oii\,3727 
is stronger than on average (TI05). 
In variant $M2cv_{\rm X}$ (Sect.~\ref{disc_stab}), where the intensity 
of \oii\ is multiplied by 1.35, both the predicted 
\feiv\,4906 and Fe/H are divided by 1.4. This lower Fe/H is adopted. 

Solar abundances are tabulated by Asplund et al. (2005, AGS05). The 
compilation by Lodders (2003) is in substantial agreement with AGS05 
($+0.03$\,dex for all O--S elements of interest here and $+0.02$\,dex for 
Fe relative to H), except for Ar/H ($+0.37$\,dex). The larger argon abundance 
is convincingly advocated by Lodders (2003). Ar/O is adopted from 
this reference. Then Ar/H coincides with the value listed by 
Anders \& Grevesse (1989). The shift of O/H from Anders \& Grevesse (1989) 
to AGS05 is $-0.27$\,dex, out of which $-$0.07\,dex corresponds to the change 
from proto-solar to solar abundances. Shifts for X/Fe are $\sim$~$-0.20$\,dex 
for N, O, Ne, $-0.11$\,dex for C, $-0.07$\,dex for S and $\sim$~0.0 for other 
elements of interest. 

In Table~\ref{tab_ab}, the present model abundances by number 12+log(X/H) 
for \IZ\,NW (`$M$') are provided in Col.\,2. The abundances X/O 
relative to oxygen from models (Col.\,3, `$M$') are compared to empirical 
values obtained by IT99 (Col.\,4, `IT'). The 
brackets [X/Y] = log(X/Y) -- log(X/Y)$_{\odot}$ from models are given 
in Cols.\,5 \& 6 (Y $\equiv$ H and Fe). Finally [X/Fe] is provided for 
Galactic Halo stars with [Fe/H]~$\sim$~$-1.8$ (Col.\,7, `H'). 
Despite considerable efforts to include 3D and non-LTE effects in 
the study of line formation in cool stars, 
astrophysical descriptions and atomic data may 
still entail uncertainties in stellar abundances 
(\eg, Fabbian et al. 2006), particularly for nitrogen. Also, a 
subpopulation of N-rich stars is well identified (\eg, Carbon et al. 1987). 

Comparing Cols.\,3 \& 4, the model and empirical X/O agree to about 0.1\,dex. 
The present C/O is close to the one obtained by Garnett et al. (1997), 
who claim that C/O is anomalously large in \IZ. IT99 argue that the 
subregion of \IZ\,NW observed with the {\sc hst} is especially hot 
according to spatially resolved MMT data and that C/O is therefore small. 
Nonetheless, IT99 also derive an exceedingly low O/H at the same 
position ``because of the higher \Te'', which poses a problem of logics 
since there is {\sl a priori} no link of causality between \Te\ and 
O/H within \IZ. The [C/O] = $-0.39$ resulting from the present model 
is indeed marginally incompatible with the up-to-date 
[C/O] = $-0.57\pm0.15$ corresponding to Galactic Halo stars with 
[O/H] = $-1.45$ (Fabbian et al. 2006). This `large' [C/O] is analysed 
by Garnett et al. (1997) in terms of carbon excess, suggesting 
that an old stellar population managed to produce this element, 
then challenging the view that \IZ\ is genuinely young 
(\eg, IT04), a view also 
challenged by Aloisi et al. (2007). From models, [C/Fe] appears 
to be identical in \IZ\ and halo stars of similar metallicity 
(Cols.\,6 \& 7). {\sl The relatively large [C/O] in \IZ\ is due to 
a relatively small [O/Fe]}. This is indirectly confirmed by the 
agreement between \IZ\ and halo stars for all elements beyond neon 
(argon should follow lighter $\alpha$-elements). The [X/O]'s 
($\equiv$ [X/H] + 1.45) are the usual basis to discuss elemental abundances 
in BCD's and nebulae. Exceptionally, in \IZ, the abundances of iron and 
heavy $\alpha$-elements are in harmony, allowing to consider the 
oxygen abundance with respect to metallicity, instead of 
defining metallicity by means of oxygen itself. Any iron locked into 
dust grains would further decrease [O/Fe] in \IZ. Apparently, for 
sufficiently low metallicity of the ISM and/or sufficient youth of 
the host galaxy, iron does not find paths to efficiently condense into dust. 
Alternatively, dust grains may be destroyed by shocks. 

\subsection{Overall evaluation of models}
\label{disc_eva}

Line intensities are generally well accounted for, although \siv\la10.5$\mu$ 
is overpredicted by 90--100\% (Sect.~\ref{disc_ir_siv}). 

\begin{table}
%Table 7  %M3 sf Si/1.5 S/1.15 Ar/1.05 Fe/1.4
\caption[]{Abundances in \IZ\,NW compared to solar and Galactic halo stars}
\begin{tabular}{l|c|cc|c|cc}
\hline\hline
   &      &     X/O & X/O    &  [X/H]  & [X/Fe] &    [X/Fe]    \\
El.& $M$$^a$&   $M$ & IT$^b$ &   $M$   &   $M$  &    H$^c$     \\
\hline
C  & 6.55 & $-$0.66 &$-$0.77 & $-$1.84 & $-$0.02&$-$0.05$\pm$0.15\\
N  & 5.60 & $-$1.61 &$-$1.56 & $-$2.18 & $-$0.36&$-$0.40$\pm$0.30\\
O  & 7.21 &    $-$  &  $-$   & $-$1.45 & \ 0.37 &\ 0.52$\pm$0.15 \\
Ne & 6.39 & $-$0.82 &$-$0.80 & $-$1.45 & \ 0.37 &    $-$     \\
Si & 5.89 & $-$1.32 &$-$1.46 & $-$1.62 & \ 0.20 &\ 0.21$\pm$0.08\\
S  & 5.57 & $-$1.64 &$-$1.55 & $-$1.57 & \ 0.25 &\ 0.21$\pm$0.07\\
Ar & 4.97 & $-$2.24 &$-$2.16 & $-$1.55 & \ 0.27 &    $-$     \\
Fe & 5.63 & $-$1.58 &$-$1.45 & $-$1.82 &   $-$  &    $-$     \\
[0.1cm]
\hline
\end{tabular}

$^a$ 12 + log(X/H) by number in present model $M$.\\
$^b$ Empirical abundance: Izotov \& Thuan (1998, 1999) with 12 + log(O/H) = 7.16; Fe: Thuan \& Izotov (2005).\\
$^c$ Halo stars [Fe/H] $\sim$ $-1.8$: Fabbian et al. (2006), Garc\'ia P\'erez et al. (2006), Nissen et al. (2007), Carbon et al. (1987).\\
\label{tab_ab}
\end{table}

The freedom left in the parameters describing the SED and the shell 
acts at the few \% level upon the \oiii\la4363 predicted intensity. Thus, 
the calculated \oiii\la4363 shifts from 96.0\% of the observed intensity 
in $M2$ ($T2$ = 4$\times$10$^{4}$\,K, $f^{cov}_i$ = 0.26, 0.30) 
to 99.6\% in $M2c$ ($f^{cov}_i$ = 0.22, 0.60) and 99.8\% in M3 
($T2$ = 5$\times$10$^{4}$\,K, $f^{cov}_i$ = 0.23, 0.50). 
Adding less than 1\% luminosity as soft X-rays 
(\eg, $M2_{\rm X}$, compared to $M2$, Sect.~\ref{disc_ir_low}), 
results in +1.7\% for \oiii\la4363. Also, increasing He/H from 
0.080 to the possibly more realistic value 0.084 (Peimbert et al. 2007), 
\oiii\la4363 is enhanced by a further +0.6\%. Since both the X-ray and He/H 
corrections are more than plausible, it is relatively easy to reach 100--102\% 
of the observed \oiii\la4363 intensity in the assumed configurations. 

However, models tell us that $r$(\oiii) can hardly be larger than the 
observed value. It may prove necessary to consider alternative gas 
distributions, \eg, in case the \siv\ misfit is confirmed by more 
accurate observational and theoretical data. Assuming larger 
densities in the diffuse medium (Sect.~\ref{disc_ir_siv}) and/or 
considering thick filaments closer to the source tend to 
penalize \oiii\la4363. 

At this point, uncertainties on \oiii\ collision strengths $\Omega$ 
need be considered (Appendix\,\ref{fiat_oiii}). Using 
$\Omega$'s by Lennon \& Burke (1994, LB94) instead of Aggarwal (1993, Ag93), 
the computed \la4363 would be 2.1\% smaller and more difficult 
to explain. On the other hand, using $\Omega$($^3$P\,$-$\,$^1$D) from Ag93 and 
$\Omega$($^3$P\,$-$\,$^1$S) from LB94 would enhance all computed \la4363 
intensities by 2.1\%. Concerning transition $^3$P\,$-$\,$^1$S which 
controls \la4363, both Ag93 and LB94 find a 5--6\% increase of $\Omega$ from 
2$\times$10$^{4}$\,K to 3$\times$10$^{4}$\,K, suggesting the influence of 
resonances. If for some reason the energy of these resonances could be 
shifted down, there would be room for a few \% increase of $\Omega$ 
at 2$\times$10$^{4}$\,K compared to the current value, then introducing 
more flexibility in the present model of \IZ. 

Summarizing, the hypothesis of pure photoionization by stars in the form 
explored here is perfectly tractable, but the models approach a limit. 
This is in a sense satisfactory, considering that \IZ\ is an extreme object 
among BCD's, but sufficient flexibility in choosing solutions is worthwile. 
An analysis of how the computed $r$(\oiii) can be influenced shows that, 
in the case of \IZ, possible variations of `astrophysical' origin are of 
the same order as the uncertainties affecting the $\Omega$'s. 
Since the set of computed $r$(\oiii) tends to be down by 2--3\% relative 
to observation, it is legitimate to question the $\Omega$'s. Now, 
the recent re-evaluation of the distance to \IZ\ by Aloisi et al. (2007) 
may offer an `astrophysical alternative': multiplying $D$, $R_{\rm i}$ and 
$R_{\rm f}$ by 2$^{1/2}$ and the luminosty by 2, the relative volume 
increase leads to smaller $P_{\rm in}$ and $\tau_{\rm c}$, and, after 
reconvergence, \oiii\la4363 is enhanced by +2.4\%. 
Noneteless, \siv\,10.5$\mu$ is enhanced too.

\section{Concluding remarks}
\label{concl}

Owing to its small heavy element content, \IZ\ stands 
at the high-\Te\ boundary of photoionized nebulae. 
Where ionization and temperature are sufficiently high, the cooling is 
little dependent on conditions, except through the concentration of H$^0$, 
controlled by density. Therefore, in the photoionization model logics, 
{\sl \Te\ is then a density indicator}, in the same way it is an O/H 
indicator in usual \hii\ regions. It is for not having recognized implications 
of this new logics, that low-metallicity BCD models failed. 

In a photoionization model study of \IZ\,NW, SS99 employed a  filling-factor 
description and concluded that \Te(\oiii) was fundamentally unaccountable.
The {\sl vogue} for this simple description of the ionized gas 
distribution resides in its apparent success for usual \hii\ regions, 
a success falling in fact to the strong dependence of cooling on abundances. 
Universally adopted along past decades, this concept 
led all authors to conclude that photoionization by hot stars did not 
provide enough energy to low-Z GEHIIRs. This conclusion is in line 
with a movement of calling into question photoionization by stars 
as the overwhelmingly dominant source of heat and ionization in 
gaseous nebulae, a movement cristalizing on the 
`$t^2$ problem' (Esteban et al. 2002; Peimbert et al. 2004), 
since the presence of \Te\ fluctuations {\sl supposedly larger} than those 
reachable assuming photoionization by stars implies additional heating. 

A conclusion of the present study is that the gas distribution is no less 
critical than the radiation source in determining the line spectrum 
of \hii\ regions. Assuming pure photoionization by stars, the remarkable 
piece of information carried by the large \Te(\oiii) of \IZ\,NW is that the 
mean density of the \oiii\ emitting region is much less than \Ne(\sii), 
a low \Ne\ confirmed by line ratios \oiv\la25.9$\mu$/\heii\la4686 and 
\feiii\la4986/\feiii\la4658. \IZ\,NW models comprising a plausible SED 
and respecting geometrical constraints can closely match almost all 
observed lines from UV to IR, including the crucial \oiii\la4363 
(\siv\la10.5$\mu$ is a factor 2 off, however). 
Thus, extra heating by, \eg, dissipation of mechanical energy in the 
photoionized gas of low-metallicity BCD galaxies like \IZ\ is 
{\sl not} required to solve the `\Te(\oiii) problem'. 
Moreover, since low-ionization fine-structure lines can be explained by 
soft X-rays, (hydrodynamical) heating is {\sl not even} required in warm 
\hi\ regions protected from ionization and heating by star radiation. 

As a final note, on close scrutiny, the solutions found here {\sl tend to be 
just marginally consistent} with observed $r$(\oiii). Given the claimed 
accuracy in the different fields of physics and astrophysics involved, 
postulating a mechanical source of heating is premature, 
whereas a {\sl 2--3\% upward correction} to the collision strength 
for transition O$^{2+}$($^3$P\,--\,$^1$S) at \Te~$\sim$~2$\times$10$^4$\,K 
is an alternative worth exploring by atomic physics. Yet, another possibility 
is a substantial increase of the distance to \IZ. Owing to accurate 
spectroscopy and peculiar conditions in \IZ, important astrophysical 
developments are at stakes in the 5\% uncertainty attached to 
\oiii\ collision strengths. 

If photoionized nebulae are shaped by shocks and other hydrodynamic effects, 
this does not imply that the emission-line intensities are 
detectably influenced by the thermal energy deposited by these processes. 
Unravelling this extra thermal energy by means of spectroscopic diagnostics 
and models is an exciting prospect, whose success depends on a 
recognition of all resources of the photoionization paradigm. 
Adopting the view that photoionization by radiation from 
young hot stars, including WR stars, is the only excitation source of 
nebular spectra in BCD galaxies, yet without undue simplifications, 
may well be a way to help progresses in the mysteries of 
stellar evolution, stellar atmosphere structure, stellar supercluster 
properties, giant \hii\ region structure and, {\sl at last}, possible 
extra sources of thermal energy in BCDs.

 % \Online


\begin{appendix} %First online appendix

\section{Models for other GEHIIRs}
\label{prevG}

Models of GEHIIRs include studies of individual 
objects and evolutionary sequences for large samples. 
Examples ordered by decreasing O/H are first reviewed. 

\subsection{Individual GEHIIRs}
\label{prev_indiv}

\subsubsection{O/H $\geq$ 1.5$\times$10$^{-4}$}
\label{prev_indiv_go}

Gonz\'alez Delgado \& P\'erez (2000) successfully model NGC\,604 
(O/H = 3$\times$10$^{-4}$) in M\,33 as a radiation-bounded sphere 
(radius 20$-$110\,pc) of density 30\cc\ and filling $\epsilon$ = 0.1, both 
\oi\la6300 and \oiii\la4363 being explained. 

Garc\'{\i}a-Vargas et al. (1997) successfully model circumnuclear 
GEHIIRs in NGC~7714 as thin, constant-density (\Nh~$\sim$~200\cc), 
radiation-bounded shells with O/H~=~(2$-$3)$\times$10$^{-4}$: $r$(\oiii) 
is accounted for within errors and \oi\ is just moderately underestimated.
The nuclear GEHIIR of NGC~7714 is modelled by Gonz\'alez Delgado et al. (1999) 
as a full sphere with very small $\epsilon$. The adopted 
O/H = 3$\times$10$^{-4}$ is too large since \oiii\la4363 is underpredicted. 
Obviously, a better fit to the available optical line spectrum could 
again be achieved for the nucleus. Gonz\'alez Delgado et al. (1999) 
are probably not well founded to invoke extra heating by shocks. 

Luridiana \& Peimbert (2001, LP01) propose a photoionization 
model for NGC~5461 (O/H = 2.5$\times$10$^{-4}$), a GEHIIR in M101. 
As for NGC~2363 (Appendix\,\ref{prev_indiv_po}), LP01 apply 
an aperture correction to their spherical model. 
The \hb\ and $r$(\sii) spatial profiles are reproduced with a Gaussian 
density distribution of very small $\epsilon$ and high inner density 
$-$ 500\cc\ compared to \Ne(\sii)\,$\sim$\,150\cc\ $-$ {\sl meant to 
increase the inner O$^+$ fraction}. In this way, the \la5007, \la4363 and 
\la3727 fluxes restricted to the theoretical slit can be 
accounted for\footnote{LP01 state {\sl a priori} that \oiii\la4363 
``almost surely has a contribution from processes other than photoionization'' 
and, consistently, conclude that their model ``fails to reproduce the 
observed \oiii\la4363 intensity'', but both statements seem to be refuted 
by pieces of evidence they present.}, but not \oi, ``a not unusual fact'', 
nor \sii, which, unlike a belief of LP01, is not enhanced by increasing the 
primary flux below 1.0\,ryd. As noted by LP01, the outputs of their model 
are strongly dependent on the density structure.

From their sophisticated study of NGC~588 (O/H = 2$\times$10$^{-4}$), 
Jamet et al. (2005, JS05) conclude that 
``the energy balance remains unexplained''. This negative conclusion 
is essentially based on the fact that \Te(\oiii), from the ratio 
\la4363\AA/\la5007\AA, is observed to be larger than \Te(\oiii, IR), 
from \la5007\AA/\la88$\mu$m, by $\Delta$$T_{obs}$ = 2\,700$\pm$700\,K, 
while the corresponding $\Delta$$T_{mod}$ is only 1\,400$\pm$200\,K in models, 
which are otherwise satisfactory, accounting reasonably well for \Te(\oiii) 
and the distribution of ionization (models DD1 and DDH exhibited by JS05, 
who carefully consider uncertainties related to the SED and 
the small-scale gas distribution). 
Considering the difficulty of calibrating the ISO-LWS fluxes relative to the 
optical and the sensitivity of the \oiii\la88$\mu$m emissivity to \Ne, 
the 400\,K gap between $\Delta$$T_{obs}$ and $\Delta$$T_{mod}$ is 
{\sl not a sound basis to claim the existence of an energy problem}.
If diagnostics based on IR lines are desirable, the energy problem 
raised so far in GEHIIR studies is not related to these lines. 
Instead of a heating problem as in \IZ, the model presented by JS05 could 
be facing a {\sl cooling} problem, since the computed \Te(\oiii, IR) 
is too high. 

An additional energy source is not clearly needed 
in the cases of NGC~588, NGC~5461, NGC~7714 and NGC~604. 

\subsubsection{O/H $<$ 1.5$\times$10$^{-4}$}
\label{prev_indiv_po}

Rela\~no et al. (2002) provide an inventory of NGC~346, a GEHIIR of 
the SMC (O/H = 1.3$\times$10$^{-4}$). Their spherical, constant-density, 
matter-bounded photoionization model, whose only free parameter is a 
filling factor $\epsilon$ ({\sl alla} SS99), accounts for the escape of 
ionizing photons, but underpredicts collisional lines, especially 
\oiii\la4363. After unsuccessful variations on geometry, the authors 
preconize, following SS99, an additional source of energy. 

After an extensive exploration of photoionization models with filling 
factor for the bright GEHIIR NGC~2363 (O/H = 8$\times$10$^{-5}$), 
Luridiana et al. (1999, LPL99) conclude that they cannot find a 
solution unless they introduce \Te\ fluctuations by hand, 
\ie, they assume a larger $t^2$ than the one intrinsic 
to their model. This $t^2$, intended to 
enhance the computed \oiii\la4363, is justified by the fact 
that the observed Paschen jump temperature is less than \Te(\oiii) 
and supported by a self-consistency argument: 
a larger $t^2$ leads to a larger O/H, hence a larger 
number of WR stars, hence (1) a larger injection of mechanical energy, 
supposed to feed the temperature fluctuations themselves, and 
(2) a larger photon flux above the He$^+$ ionization limit, useful to 
increase \heii\la4686. However, as acknowledged by Luridiana et al. (2001), 
the WR star winds make a poor job to generate a significant $t^2$ in NGC~2363. 
Also, present views suggest that arguments based on WR stars in low-Z 
galaxies were illusory just a few years ago (\eg, Leitherer 2006; 
Appendix\,\ref{WRstars}). Finally, $r$(\oiii), underestimated by only 
$\sim$~12\% in the `standard' low-Z model by LPL99, is divided by 2 on 
using the larger O/H, so that a relatively minor difficulty is first made 
much worse and then solved by means of a $t^2$. The slit correction 
advocated by LPL99 is considered in Appendix\,\ref{sphere_slit}. 

Luridiana et al. (2003, LPPC03) consider a spherical model for a GEHIIR 
of SBS~0335-052 (O/H = 2$\times$10$^{-5}$). A Gaussian distribution 
with large maximum density and small $\epsilon$ proves unsatisfactory. 
LPPC03 then consider a 10-shell model (over 50 free parameters, 
most of which pre-defined), in which {\sl each shell is radiation 
bounded} and is characterized by a covering factor. Although each 
shell is still given an $\epsilon$, the new model is equivalent 
to a collection of geometrically thin radiation-bounded sectors 
at different distances from the source (see also Giammanco et al. 2004) 
and ``gracefully reproduces the 
constancy of the ionization degree along the diameter of the nebula''. 
Hence, the authors are forced by observational evidences 
to {\sl implicitly abandon} the classical filling-factor approach. 
Nonetheless, whatever the complexity of these models, all of them 
fail to account for the high \Te(\oiii).

LPPC03 consider a Gaussian model for \IZ\,SE (O/H = 1.7$\times$10$^{-5}$) 
with again a relatively large maximum density and, unlike for SBS~0335-052, 
a relatively large $\epsilon$, resulting in a rather compact model nebula, 
in which the computed \Te(\oiii) compares quite well with the observed one. 
Unlike for the NW, the {\sc hst} image (Cannon et al. 2002) of the younger 
SE \hii\ region does not show a shell surrounding an MSC.
Nonetheless, considering the strong output of mechanical energy from 
massive stars, it is likely that inner cavities already developped. 
The strong indirect evidence for too compact a gas distribution in the 
model by LPPC03 is the notable weakness of the computed intensity of 
\oii\ and other low-ionization lines. Adopting a more expanded structure in 
order to increase \oii, yet keeping the general trend of the gas distribution, 
the computed \Te(\oiii) would be forseeably lower than in the model by LPPC03. 

\subsection{Individual GEHIIRs: discussion}
\label{prev_indiv_disc}

LPPC03 describe the `\Te(\oiii) problem' they face in their study of 
SBS\,0335-052 (Appendix\,\ref{prev_indiv_po}) as ``a systematic feature'' 
of \hii\ region models and, following SS99, they state that 
this problem ``can be ascribed to an additional energy source acting in 
photoionization regions, other than photoionization itself''. Nevertheless, 
\Te(\oiii) seems to be accountable in existing photoionization models for 
GEHIIRs with, say, O/H $\geq$ 1.5$\times$10$^{-4}$
(Appendix\,\ref{prev_indiv_go}). Similarly, the computed 
\oiii\la4363 is correct, possibly even too large, for \hii\ regions 
of the LMC (Oey et al. 2000). If, despite apparent complementarities, 
the \Te(\oiii) and $t^2$ problems have different origins (Sect.~\ref{intro}), 
no `systematic feature' can be invoked.  

In modelling objects with near solar 
abundances, \oiii\la4363 is controlled by O/H, \oiii\la5007 by the `color 
temperature' of the ionizing radiation, \oii\la3727 by the ionization 
parameter, while \oi\ is maximized in radiation-bounded conditions. For 
these objects, assuming a `large' (constant) density, \eg, $\sim$~\Ne(\sii), 
associated to an {\sl ad hoc} $\epsilon\ll$~1, is often successful, 
{\sl although this does not prejudge of the relevance of the model found}. 
Indeed, this assumption proves to be at the heart of the \Te(\oiii) problem 
met in low-Z BCDs (Appendix\,\ref{filling}). 

\subsection{Extensive analyses of BCDs}
\label{prev_extens}

The conclusion of an early extensive analysis based on radiation-bounded, 
low-density full sphere models for low-Z BCDs (Stasi\'nska \& Leitherer 1996) 
is optimistic concerning \Te(\oiii), whereas \oi\ is then qualitatively 
explained in terms of shock heating. Nevertheless, in an extension of this 
study to large-Z objects with no measured \Te(\oiii), 
Stasi\'nska et al. (2001) reinforce the energy problem 
raised by SS99 (Sect.~\ref{prev_oiii}) when they conclude that ``a purely 
`stellar' solution seems now clearly excluded for the problem of 
\oiii/\hb\ {\sl versus} \oii/\hb\  as well as \sii/\hb'', while, conversely, 
they still endorse the unproved statement of SS99 (Sect.~\ref{prev_oi}) that 
``strong \oi\ emission is easily produced by photoionization models in dense 
filaments''. 

The sequence of 
photoionization models proposed by Stasi\'nska \& Izotov (2003, SI03) 
for a large sample of low-Z BCDs (divided in three abundance bins) 
illustrates views expressed after the failure of models acknowledged by SS99 
for \IZ\ (Sect.~\ref{prev_IZ}). In the description by SI03, 
an evolving synthetic stellar cluster (10$^5$\Msun, 
instantaneous burst) photoionizes a spherical shell of constant density 
\Nh\ = 10$^2$\cc\ at the boundary of an adiabatically expanding hot bubble. 
With suitable bubble properties, underlying old stellar population, aperture 
correction and time evolution of the covering factor, the range of \hb\ 
equivalent width (EW(\hb)) and the trends of \oiii\,5007, \oii\,3727, 
\oi\,6300 {\sl versus} EW(\hb) can be reproduced for the high-Z bin 
(O/H $\sim$ 1.5$\times$10$^{-4}$) within the scatter of the data. 

Applying similar prescriptions to the intermediate-Z bin, 
the oxygen lines and \heii\la4686 (\heii\ was just 
fair in the first bin) are underpredicted. 
SI03 diagnose an insufficient average energy per absorbed photon 
and assume that the stellar cluster is supplemented by a 
{\sl strong 10$^6$\,K bremsstrahlung-like radiation source}, which 
solves the \heii\ problem (He$^+$ is further ionized by extra 4--5\,ryd 
photons; see, however, Appendix\,\ref{WRstars}) and alleviates the \oi\ 
problem (the soft X-rays further heat and widen the ionization front), 
but barely improves \oiii\ and \oii. Agreement of the model sequence 
with observation is finally restored by supposing {\sl in addition} 
that the shell includes a {\sl time-variable oxygen-rich gas component} 
attributed to self-enrichment: in the example shown by SI03, this component 
is 4-fold enriched in CNO, etc. relative to the original abundance and 
encompasses half of the shell mass after a few \My, so that one generation 
of stars produced a 2.5-fold enhancement of the average abundance in the 
photoionized gas. 

This description essentially applies to the low metallicity bin 
(O/H $\sim$ 2$\times$10$^{-5}$) of particular concern for \IZ, but with 
{\sl even more extreme properties} for the O-rich component, since it 
should be overabundant by 1\dx, resulting in a {\sl 5-fold enhancement 
of the final average abundance}. 

\subsection{Extensive analyses of BCDs: discussion}
\label{prev_extens_disc}

The time scale of 0.5\My\  for the growth of the O-rich 
component in the description by SI03 cannot directly fit in the 
self-pollution scenario since it is shorter than the stellar evolution 
time scale. Also, a sudden {\sl oxygen} self-pollution of the gas is not 
observed in supernova remnants. 

The assumed X-ray power is $\sim$\,10\% of the cluster luminosity or 
$\sim$\,2\dx\ times the estimated X-ray {\sc rosat} power (0.07$-$2.4 keV) 
of the hot bubble fed by stellar winds and supernovae around a usual 
MSC (Strickland \& Stevens 1999; Cervi\~no et al. 2002). 
Moreover, the hot gas is generally raised at {\sl several} 10$^6$\,K 
(Stevens \& Strickland 1998). Adopting a larger temperature, the X-ray power 
should be even larger, as only the softer radiation interacts usefully with 
the ionized gas. \IZ\ itself is a rather strong X-ray emitter in the 
0.5--10 keV range, yet 20 times weaker than the source assumed by SI03 
(Thuan et al. 2004; Sect.~\ref{disc_ir_low}). 

Apart from these problems, SI03 do not address the question of the intensity 
of \oiii\la4363. The narrow radiation-bounded shell adopted by SI03 usefully 
favours \oi\la6300, but makes the computed intensity 
of \oiii\la4363 even worse than the one obtained by, \eg, SS99 
(Sect.~\ref{prev_IZ}). Moreover, adding the prominent O-rich component 
advocated by SI03 will (1) decrease \oiii\la4363 by a further 30$-$40\% 
on average and (2) conflict with the existence of very low-Z BCDs, 
since any of them will be condamned to shift to the intermediate class 
defined by SI03 after just 1 or 2\My.

If what SI03 qualify as ``appealing explanations'' presents any 
character of necessity for BCD models, then the hypothesis of photoionization 
by stars, which was already given a rough handling by SS99 in their analysis 
of \IZ, should be considered as definitively burried for the whole 
class of low-Z BCDs. The fact that SI03 discard \oiii\la4363 in 
their analysis confirms that they endorse and reinforce views 
expressed by SS99 or Stasi\'nska et al. (2001) and give up explaining 
\Te(\oiii) in low-metallicity GEHIIRs by means of stellar radiation. However, 
{\sl the same restrictive assumption} as for individual GEHIIR 
studies (Appendices\,\ref{prev_indiv}--\ref{prev_indiv_disc}) bears on the 
gas distribution adopted by SI03, since their (geometrically thin) `high' 
constant-\Nh\ model sphere is nothing but a zero-order approximation to 
a model shell with classical filling factor (Appendix\,\ref{filling}). 

\end{appendix}

\begin{appendix} %Second online appendix

\section{On the gas distribution in GEHIIRs}
\label{gas_distrib}

\subsection{Misadventures of the filling factor concept}
\label{filling}

For the sake of reproducing the \ha\ surface brightness of \hii\ 
regions, asssuming a gas density much larger than $<$\Ne$^2$$>$$^{1/2}$, 
the `filling factor paradigm' posits that the emitting gas belongs to 
{\sl optically thin}, `infinitesimal' clumps, filling altogether a 
fraction $\epsilon\ll 1$ of the volume.

Given that the 
stellar evolution timescale exceeds the sound crossing time of \hii\ 
regions, small optically thin ionized clumps will have time to expand 
and merge into finite-size structures. {\sl If} these structures have 
the original density, they are likely to have {\sl finite or large 
optical depths, in contradiction with the filling factor concept}. 

Filamentary structures, ubiquitous in \ha\ images of nearby GEHIIRs, are often 
taken as justifications for introducing $\epsilon$ in photoionization models. 
However, (1) the geometrical thickness of observed filaments is consistent 
with radiation-bounded structures and (2) an individual filament most often 
emits both high and low ionization lines (\eg, Tsamis \& P\'equignot 2005). 
{\sl The filling-factor description is flawed}. 

A GEHIIR may well be a collection of {\sl radiation-bounded} 
filaments embedded in coronal {\sl and} photoionized diffuse media. 
The idea behind assuming this configuration is that only the ionized 
`atmospheres' of long-lived, radiation-bounded, evaporating structures 
will possibly maintain a substantial overpressure relative to 
their surroundings (a similar idea applies to the ``proplyds'' 
found in Orion; \eg, Henney \& O'Dell 1999). 

The filling factor concept fails on both theoretical and observational 
grounds. Nevertheless, introduced as a technical tool to manage \Ne\ 
diagnostics like $r$(\sii), $\epsilon$ came to be improperly used to adjust 
the {\sl local} ionization equilibrium of the gas through \Nh, in an effort 
to overcome problems of ion stratification generated by the filling factor 
description itself (Appendix\,\ref{sphere_slit}). 

The \Te(\oiii) problem met in oxygen-poor GEHIIRs may relate to the 
{\sl loss of plasticity} affecting photoionization models, as the dependence 
of gas cooling on abundances vanishes. Then, {\sl cooling} depends on 
the relative concentration of H$^0$ (collisional excitation of \lya), 
{\sl controlled by the local \Nh}. Hence, the (improper) freedom on 
$\epsilon$ is eroded. Moreover, if density is not uniform, 
\Ne(\sii) is a biased estimate for \Ne\ in the bulk 
of the emitting gas, since S$^+$ ions will belong to dense, 
optically thicker clumps. Emission from an interclump medium with 
\Ne\,$<$\Ne(\sii) will selectively enhance the computed \oiii\la4363 
intensity. LPL99 state promisingly that their model includes 
``denser condensations uniformly distributed in a more tenuous gas'', 
but in practice only the condensations emit. This restriction is shared 
by virtually all published models for low-Z GEHIIRs. While the assumed 
density of the emitting gas can be orders of magnitude larger than 
$<$\Ne$^2$$>$$^{1/2}$, emission from a lower density gas is neglected 
{\sl by construction} (The study by JS05 is an exception, 
but NGC~588 is not low-Z; Appendix\,\ref{prev_indiv_go}). 
The \Te(\oiii) problem suggests lifting this restriction.

\subsection{Spheres, slits, filling factor and stratification}
\label{sphere_slit}

Spherical models raise the question of how to 
compare computed spectra with nebular spectra observed
through, \eg, a narrow slit. 
LPL99 advocate extracting emission from 
that part of the sphere which would project on the slit. 
Despite obvious problems with non-sphericity, LPL99 
and others argue that this procedure would at least allow weighting 
the contributions from low- and high-ionization zones in a more realistic 
manner. Using the classical filling factor concept (Appendix\,\ref{filling}) 
in GEHIIR models, ion stratification spreads over the 
whole nebula and the \oi\ emission is effectively confined to 
outer layers, in which the primary radiation eventually vanishes. 
If, on the contrary, the emitting gas belongs to {\sl radiation-bounded} 
filaments distributed within the nebula, then {\sl ion stratification 
disappears to first order}. Radial ionization gradients, if any, are 
no more related to a progressive destruction of primary photons along 
the full radial extension of the nebula, but to changes in (local) average 
ionization parameter. 

LPL99 conclude that \oi\ is due to shock excitation in NGC~2363 
(Sect.~\ref{prev_indiv_po}) because the computed intensity is weak in their 
theoretical slit extraction. Nonetheless, the \oi\ intensity is fairly 
correct in their global spectrum. This apparent failure of their 
photoionization model may well be due to the unfortunate combination of 
(1) a very small $\epsilon$ and 
(2) the extraction of a slit shorter than the diameter of the model sphere. 
Along the same line, LPPC03 are confronted to undesirable consequences 
of the filling factor assumption on the variation of ionization along a 
slit crossing SBS~0335-052 (Appendix\,\ref{prev_indiv_po}). 

If a geometrically defined model can hardly provide an approximation to a 
complex \hii\ region, thus casting doubts on theoretical slit extractions, 
{\sl global spectra} are less sensitive to geometry, owing to 
conservation laws. 

Moreover, in computing 1-D photoionization models, the (spherical) symmetry 
enters {\sl only} in the treatment of the diffuse ionizing radiation field, 
which is generally not dominant in the total field. The diffuse 
field, most effective just above the ionization limits of H, He and He$^+$, 
is relatively {\sl local} at these photon energies (in accordance with 
the `Case~B' approximation) and little dependent on global geometry. Let us 
define an ``elementary spherical model'' (for given SED) by a radial density 
distribution of whatever complexity. Since the local state of the 
gas is chiefly related to the primary (radial) radiation, 
a composite model made of a judicious 
combination of elementary spherical models, each of them restricted 
to a sector characterized by a {\sl covering factor}, can provide 
{\sl topologically significant and numerically accurate} descriptions of 
global spectra for nebulae with complex structures. Defining a 
`topology' as a particular set of spherical models with their attached 
covering factors, any given topology is in one-to-one correspondence with 
a global spectrum {\sl and a full class of geometries}, since any sector 
can be replaced by an arbitrary set of subsectors, provided that 
the sum of the covering factors of these subsectors is conserved.

Thus, a good modelling strategy for a GEHIIR is one in which a global 
(probably composite) model spectrum is compared to the observed global 
spectrum. If only one slit observation is available, given that the ion 
stratification tends to be relatively loose and erratic in GEHIIRs, 
it is wise to directly use this spectrum as the average spectrum 
(together with scaling by the absolute \hb\ flux), with the 
understanding that the resulting photoionization model will 
represent a `weighted average' of the real object. For many 
practical purposes, this weighting may not significantly impact 
on the inferences made from the model, unless the slit position is 
exceedingly unrepresentative. 

\end{appendix}

\begin{appendix}

\section{\heii\la4686, WR stars and SEDs}
\label{WRstars}

\IZ\ harbours Wolf-Rayet (WR) stars (Legrand et al. 1997; Izotov et al. 1997; 
de Mello et al. 1998; Brown et al. 2002). WR stars have been challenged 
as the sole/main cause of nebular 
\heii\la4686 in BCDs on the basis of a lack of correlation between 
the occurence of this line and the broad `WR bumps' (\eg, Guseva et al. 2000). 
The study of WR stars is experiencing a revolution 
(Maeder et al. 2005; Meynet \& Maeder 2005; Gr\"afener \& Hamann 2005; 
Vink \& de Koter 2005; 
Crowther 2007) after the realization that (1) {\sl rotation} of massive 
stars favours enhanced equatorial mass loss, element mixing by shears, 
and angular momentum transport by meridian circulation, 
(2) low-Z massive stars tend to be fast rotators and accelerate as they 
evolve off the main sequence, so that the lower mass limit for a star to 
become a WNE star is much reduced, and (3) for a given type of WR star, 
the mass loss is lower for lower metallicity (Fe/H, not O/H), with three 
consequences: the broad WR features are less evident for low metallicity 
(weaker optical continuum {\sl and} smaller EW of WR bumps), 
the duration of the WR stage can be longer, and the EUV luminosity is larger 
due to reduced blanketing effect. Thus, the above lack of correlation 
can now be partly 
ascribed to a {\sl bias}, related to the tendency of WR star atmospheres to 
display less prominent optical signature when they emit more EUV radiation. 
{\sl The WR star population of \IZ\ and the ability of 
these stars to emit radiation beyond 4\,ryd have almost certainly 
been grossly underestimated} (Crowther \& Hadfield 2006). 

Other observations, \eg, 
for SBS 0335 052E (Izotov et al. 2001b; Izotov et al. 2006b) 
are still taken as evidence for \heii\ excitation by radiation 
from very fast shocks: (1) the \heii\ line is broader than other nebular 
lines, (2) the \heii\ emission is spread out far away from 
the main MSCs, and (3) \Te\ is larger in \heii\ emitting area, hence at 
large distances from the main ionizing sources. These findings 
are definitely {\sl no} compelling arguments against photoionization by 
WR stars. The larger \heii\ line width indicates larger turbulence and/or 
velocity gradients, not necessarily shocks. That \Te\ is observed to be 
larger in \heii\ emitting gas is in agreement with photoionization models. 
The spatial extent of \heii\ may reflect the distribution 
of a few WR stars, which may not belong to the main cluster 
and may not be easily detected (Crowther \& Hadfield 2006). 
Alternatively, \heii\ can be produced far from the ionizing stars if 
the medium is porous and permeated by low density, optically thin gas, 
\eg, along a galactic wind outflow (Izotov et al. 2006b). The picture of 
a galactic wind also suggests an explanation for the \heii\ width. 

Photoionization models are test beds for ionizing radiation sources, but 
inferences on the physics of GEHIIRs should not depend on uncertain SEDs. 
Existing synthetic star clusters are inadequate to model \IZ. 
Apart from known problems with star sampling 
(Cervi\~no et al. 2003; Cervi\~no \& Luridiana 2006), 
limited knowledge of the history of actual MSCs and current uncertainties 
about WR stars, new free parameters (initial 
angular momentum and magnetic field of individual stars; rate of 
binarity) will broaden the range of possible SED evolutions, while collective 
effects in a compact cluster of massive stars may influence the output of 
ionizing radiation far from it, due to high-density stellar winds 
(Thompson et al. 2006). 

These comments justify (1) the assumption of an excitation of \heii\ 
solely by WR stars and (2) the use of a flexible analytical SED for 
\IZ\,NW (Sect.~\ref{mod_star}). 

\end{appendix}

\begin{appendix}

\section{Atomic data}
\label{fiat}

\subsection{Collisional excitation of \hi}
\label{fiat_h}

Collision strengths $\Omega$($1s$--$nl$) ($n<6; l<n$) for \hi\ are taken 
from Anderson et al. (2000, ABBS00). The $\Omega$'s for $1s$--$2s$ and 
$1s$--$2p$ are much larger than for the next transitions $1s$--$nl$ and are 
not controversial. The main coolent agent in low-Z BCDs should be correctly 
implemented in all codes. Nonetheless, in the conditions of \IZ, the results 
for transitions 1--2 by ABBS00 are about 10\% larger than those carefully 
fitted by Callaway (1994), giving an estimate of possible uncertainties. 
The adopted data tend to enhance the cooling with respect to earlier data and 
to (conservatively) worsen the `\Te(\oiii) problem'. Total $\Omega$(1--$n$)'s 
listed by Przybilla \& Butler (2004) virtually coincide with ABBS00 
values for 1--2, confirming the \hi\ cooling rate, but diverge from ABBS00 
for $n>2$ and increasing \Te\ similarly to early, probably wrong, data 
(see P\'equignot \& Tsamis, 2005). 

\begin{table}
%Table App B
\caption[]{Effective collision strengths for \oiii}
\begin{tabular}{l|llll}
\hline\hline
\Te/10$^4$\,K: &    0.5 &    1.0 &    2.0 &   3.0 \\ 
\hline
Reference:$^a$ & \multicolumn{4}{c}{$\Omega$($^3$P\,$-$\,$^1$D)} \\
[0.1cm]
Sea58         & -      & 1.59   & -      & -      \\      
SSS69         & -      & 2.39   & -      & -      \\  
ENS69         & 1.85   & 2.50   & 2.91   & 2.96   \\       
ES74          & 2.17   & 2.36   & 2.55   & -      \\   
Men83         & 2.02   & 2.17   & 2.39   & -      \\
Ag83         & 2.035  & 2.184  & 2.404  & 2.511  \\
BLS89         & 2.10   & 2.29   & 2.51   & 2.60   \\
Ag93         & 2.039  & 2.191  & 2.414  & 2.519  \\
LB94          & 2.1268 & 2.2892 & 2.5174 & 2.6190 \\ 
AgK99$^b$     & 2.0385 & 2.1906 & 2.4147 & 2.5191 \\
[0.1cm]
LB94/AgK99$^c$& 1.0435 & 1.0450 & 1.0425 & 1.0397 \\
[0.1cm]
       & \multicolumn{4}{c}{$\Omega$($^3$P\,$-$\,$^1$S)} \\
[0.1cm]
Sea58         &   -    & 0.220  &   -    &  -     \\     
SSS69         &   -    & 0.335  &   -    &  -     \\    
ENS69         & 0.255  & 0.298  & 0.331  & 0.339  \\     
ES74          & 0.276  & 0.325  & 0.356  &    -   \\    
Men83         & 0.248  & 0.276  & 0.314  &    -   \\    
Ag83         & 0.2521 & 0.2793 & 0.3162 & 0.3315 \\   
BLS89         & 0.260  & 0.287  & 0.318  & 0.331  \\   
Ag93         & 0.2732 & 0.2885 & 0.3221 & 0.3404 \\   
LB94          & 0.2720 & 0.2925 & 0.3290 & 0.3466 \\   
AgK99$^b$     & 0.2732 & 0.2885 & 0.3221 & 0.3404 \\
[0.1cm]
LB94/AgK99$^c$& 0.9956 & 1.0139 & 1.0214 & 1.0182 \\
[0.1cm]
              & \multicolumn{4}{c}{$\Omega$($^1$D\,$-$\,$^1$S)} \\
[0.1cm]
Sea58         &   -    & 0.640  &   -    &   -    \\     
SSS69         &   -    & 0.310  &   -    &   -    \\  
ENS69         & 0.483  & 0.578  & 0.555  & 0.510  \\   
ES74          & 0.807  & 0.856  & 0.752  &   -    \\   
Men83         & 0.516  & 0.617  & 0.634  &   -    \\   
Ag83         & 0.5463 & 0.6468 & 0.6670 & 0.6524 \\   
BLS89         & 0.59   & 0.677  & 0.664  & 0.634  \\  
Ag93         & 0.4312 & 0.5227 & 0.5769 & 0.5812 \\   
LB94          & 0.4942 & 0.5815 & 0.6105 & 0.6044 \\   
AgK99$^b$     & 0.4312 & 0.5227 & 0.5769 & 0.5812 \\
[0.1cm]
LB94/AgK99$^c$& 1.1461 & 1.1125 & 1.0582 & 1.0399 \\
[0.1cm]
\hline
\end{tabular}

\ \ $^a$ Refs: 
Sea58: Seaton(1958); 
SSS69: Saraph et al. (1969); 
ENS69: Eissner et al. (1969); 
ES74:  Eissner \& Seaton (1974); 
Men83: Mendoza (1983); 
Ag83: Aggarwal (1983); 
BLS89: Burke et al. (1989); 
Ag93: Aggarwal (1993); 
LB94:  Lennon \& Burke (1994); 
AgK99: Aggarwal \& Keenan (1999).\\
\ \ $^b$ Results from Aggarwal (1993).\\
\ \ $^c$ Collision strength ratio.\\
 
\label{tab_oiii}
\end{table}

\subsection{Collisional excitation of \oiii}
\label{fiat_oiii}

Effective collision strengths $\Omega$ obtained over past 50 years are 
listed in Table~\ref{tab_oiii} at four \Te's for 
transitions $^3$P\,$-$\,$^1$D, $^3$P\,$-$\,$^1$S and $^1$D\,$-$\,$^1$S. 
Aggarwal \& Keenan (1999) did not feel it necessary to update earlier values 
by Aggarwal (1993; Ag93), almost contemporary with Lennon \& Burke (1994). 
The ratios of the recent values are given in Table~\ref{tab_oiii}. 
The differences are over 4\% for $^3$P\,$-$\,$^1$D 
and 10\% for $^1$D\,$-$\,$^1$S (6\% in \IZ\ conditions), but 
the latter has no influence at low \Ne. 
{\sc nebu} includes a fit better than 0.5\% to Ag93 data. 

The \oiii\ transition probabilities used in {\sc nebu} are from 
Galavis et al. (1997, GMZ97). The accuracy of the Opacity Project (OP) 
data for these transitions is 8$-$10\% (Wiese et al. 1996). 
Coherently, the much more elaborate results by GMZ97 differ 
from the OP results by 9.6\% and 5.5\% for A($^1$D\,$-$\,$^1$S) 
and A($^1$P\,$-$\,$^1$S) respectively. Would A($^1$D\,$-$\,$^1$S) 
change by as much as 5\%, the branching ratio of \oiii\la4363 
would change by 0.6\%. 
%A($^1$D\,$-$\,$^1$S)/(A($^1$D\,$-$\,$^1$S)+A($^3$P\,$-$\,$^1$S)) = 0.8731, 

Thus, discrepancies not exceeding 5\% exist among different calculations 
(3\% for $\Omega$ {\sl ratio}s), suggesting that uncertainties on the 
computed $r$(\oiii) are probably $<$~5\%. The 25--30\% 
underestimation found by SS99 is not due to erroneous atomic data. 

\subsection{Miscellaneous data}
\label{fiat_misc}

The adopted table for radiative and dielectronic recombinations 
is limited to the 11 sequences H-like$-$Na-like (Badnell 2006). Dielectronic 
rates for \sii$-$\siv\ are obtained by Badnell (1991), but total recombination 
coefficients for (recombined ions) \sid, \sii, \siii, \arv, \feii$-$\fev, 
are taken from Nahar and co-workers (Nahar, 2000 and references cited). 
The recombination rate for \sii\ used in this and previous {\sc nebu} 
computations is 1.15 times the Nahar's value. Empirical total rate 
coefficients based on PN models (P\'equignot, unpublished), implemented 
in {\sc nebu} for a decade, are 5 and 8 times the radiative ones 
for \arii\ and \ariii\ respectively. A larger factor is suspected 
for \ariii\ at high \Te. 

Collision strengths of special mention include those for \oii\ 
(Pradhan et al. 2006; also Tayal 2006b), \oiv\ (Tayal 2006), 
\siii\ (Tayal \& Gupta 1999), \siv\ (Tayal 2000) and 
\fev\ (W\"oste et al. 2002). Collisions with H$^0$ are considered 
in Sect.~\ref{disc_ir_low}. Charge exchange rates with H$^0$ for 
O$^{2+}$ and N$^{2+}$ are now from Barrag\'an et al. (2006). 

\end{appendix}

\end{document}